\documentclass[notitlepage,a4paper,fleqn,11pt]{article}
\usepackage{latexsym}
\usepackage[dvips]{color}
\usepackage{graphicx}
\usepackage{amssymb}
\usepackage[top=1.25in, bottom=1.25in, left=1.35in, right=1.35in]{geometry}
\usepackage{amsmath}
\usepackage{graphicx}
\usepackage{float}
\usepackage{parskip}
\usepackage{textcomp}
\usepackage[font=footnotesize,labelfont=bf]{caption}
\usepackage{subfig}
\usepackage{array}
\usepackage{longtable}

\numberwithin{equation}{section}
\numberwithin{figure}{section}
\title{{Nodal domains of the equilateral triangle billiard}}
\author{Rhine Samajdar$^{1}$ and Sudhir R. Jain$^{2}$\\
$^{1}${\it Indian Institute of Science, Bangalore 560012, India.}\\
$^{2}${\it Nuclear Physics Division, Bhabha Atomic Research Centre, Mumbai 400085, India.}}
\date{}
\begin{document}
\maketitle

\begin{abstract}

\noindent
We characterise the eigenfunctions of an equilateral triangle billiard in terms of its nodal domains. The number of nodal domains has a quadratic form in terms of the quantum numbers, with a non-trivial number-theoretic factor. The patterns of the eigenfunctions follow a group-theoretic connection in a way that makes them predictable as one goes from one state to another. Extensive numerical investigations bring out the distribution functions of the mode number and signed areas. The statistics of the boundary intersections is also treated analytically. Finally, the distribution functions of the nodal loop count and the nodal counting function are shown to contain information about the classical periodic orbits using the semiclassical trace formula. We believe that the results belong generically to non-separable systems, thus extending the previous works which are concentrated on separable and chaotic systems.  

\end{abstract}
\setlength{\tabcolsep}{16.1pt}
\renewcommand{\arraystretch}{2.5}

\section{Introduction}
\noindent
The experimental study of vibrating plates begins with Chladni in 1787 \cite{wood}. By sprinkling fine sand on a metallic disc supported at its centre, the act of drawing a bowstring across an edge made sand particles to jump into a geometrical design. Chladni demonstrated that the pattern thus formed was the nodal pattern of the plate - he experimented with square, circular, hexagonal, rectangular, elliptical, semicircular, and triangular plates. From that time, the subject has attracted a lot of attention of mathematicians and physicists \cite{courant}. 

The questions related to the quantification of nodal patterns have persisted through the transition from classical to quantum physics \cite{ss}. Counting and classifying patterns is quite difficult in general. A lot of progress has been made in recent times, mostly when the plates are separable or fully chaotic as we describe below. In present times, ``Chladni experiments" have turned into studies on billiards wherein a point particle moves freely inside an enclosure, reflecting specularly from the boundaries in accordance with the  Snell's law. The quantum problem then involves solving the Schr\"{o}dinger equation for the system with Dirichlet boundary conditions. The eigenfunctions obtained there are then subjected to the study of nodal lines and domains. 

\noindent
The nodal domains of a real wavefunction may be defined as  maximally connected regions wherein the function does not change sign. The sequence of the
number of nodal domains of the eigenfunctions of the Schr\"{o}dinger equation contains significant geometric information about the system \cite{toth2}. As shown in \cite{uzy2002}, the distribution of nodal domains can serve to
distinguish between systems which are integrable (with the Laplacian being separable) and non-integrable (with classically chaotic ray dynamics), and the limiting distribution, which always exists for
separable domains, can give a new criterion for chaos in quantum mechanics. The problem of the equilateral triangle billiards presents an example of a non-separable system that is analysed here. We present our study  on this system by making extensive numerical investigations. Analysis through this enormous data leads us to several interesting conjectures (Section 2), connections with the symmetry group and statistical studies on domains (Section 3) where some analytical results are also given. 

It may be worth pointing out that the number of nodal domains for a separable billiard like a rectangle is quite simple. This is because the eigenfunctions are products of trigonometric functions (sine functions) in $x$ and $y$ variables. This results in the total number of domains equalling the product of the quantum numbers. The complication arises from the fact that the ordering of energy levels (labelled by the index $j$) does not follow the number of nodal domains, $\nu _j$. Thus the distribution of the mode  number $\nu _j/j$ presents a difficult problem. There exists a limiting distribution for this system, found in \cite{shankar}. For the right isosceles triangle, a non-separable system, the eigenfunctions can be written as a sum of two terms, each a product of the type one has in a rectangular case. For this case, some inspired guess work and graph-theoretic methods help in getting some interesting results \cite{aronovitch}. In the case of an equilateral triangle, there are three terms and, as mentioned above, as one considers even slightly higher lying states,  it becomes clear that the graph-theoretic method becomes hard to apply \cite{sravan}.  
\\
\\
Let ${\cal D} \subset \mathbb{R}^2$ be the equilateral triangle of area $\displaystyle {\cal{A}} = \frac{\sqrt{3}\pi^2}{4}$ represented as 
\begin{eqnarray}
{\cal D} &=& \bigg\{(x,y) \in \bigg[0,\frac{\pi}{2}\bigg] \times \bigg[0,\frac{\sqrt{3}\pi}{2}\bigg] : y\le\sqrt{3}x\bigg\} 
\nonumber \\&\cup& \bigg\{(x,y) \in \bigg[\frac{\pi}{2},\pi \bigg] \times \bigg[0,\frac{\sqrt{3}\pi}{2}\bigg] : y\le\sqrt{3}(\pi-x)\bigg\}. 
\end{eqnarray}
\noindent The eigenvalue problem, which is defined by the Laplacian $\Delta$ on a compact Riemannian manifold of dimension two along with the imposition of Dirichlet boundary conditions on the manifold, is
stated as
\begin{equation}
-\Delta  \psi(x,y) = -(\partial^{2}_{x} + \partial^{2}_{y})   \psi(x,y) = E  \psi(x,y) \mbox{ and } \psi(x,y)|_{\partial {\cal D}} = 0.
\end{equation}

\noindent The eigenfunctions of the Laplace-Beltrami operator for the general equilateral triangle billiard, with $L$ being the length of each side, form a complete orthogonal basis and are given
by \cite{brack, McCartin} as
\begin{eqnarray}
\label{eq: wf}
\psi_{m,n}^{c,s} (x,y) &=& (\cos,\sin)\bigg[ (2m-n)\frac{2\pi}{3L}x \bigg] \sin{\bigg(n\frac{2\pi}{\sqrt{3}L}y\bigg)} 
\nonumber \\&-&   (\cos,\sin)\bigg[(2n-m)\frac{2\pi}{3L}x\bigg]\sin{\bigg(m\frac{2\pi}{\sqrt{3}L}y\bigg)} 
\nonumber \\&+&  (\cos,\sin)\bigg[-(m+n)\frac{2\pi}{3L}x\bigg]\sin{\bigg[(m-n)\frac{2\pi}{\sqrt{3}L}y\bigg]}, 
\end{eqnarray}
where $m$ and $n$ are integer quantum numbers with the restriction $m\geq2n$ and $m,n>0$. The eigenfunctions $\psi_{m,n}^c$ and $\psi_{m,n}^s$ correspond to the symmetric and antisymmetric modes respectively \cite{McCartin}. The spectrum of the eigenvalues of the Hamiltonian for the system is given by
\\
\begin{eqnarray*}
E_{m,n} = \frac{16}{9}\frac{\pi^2\hbar^2}{2\mu L^2}(m^2+n^2-mn), \mbox{ where $\mu$ is the mass of the particle.}
\end{eqnarray*}
\\
\noindent
This spectrum is arranged in an increasing order of the energy eigenvalues and is represented as the sequence $\{E_j\}^\infty_{j=1}$ such that $E_j \le E_{j+1}$. The number of nodal domains for
$\psi_{m,n}$ is denoted by $\nu_{m,n}$ and $\{\nu_j\}^\infty_{j=1}$ thus defines the sequence of nodal domain counts. Certain degeneracies inevitably arise in the spectrum due to the fact that $E_{m,n}$ is the
same for both $\psi_{m,n}^c$ and $\psi_{m,n}^s$ \cite{itzykson}. For such cases, the wavefunction for the anti-symmetric mode is assigned an index $ j_i+1$, where $j_i$ is the index of the corresponding symmetric mode.
This is justified by the consideration that the ground state, which has no nodes and thus must be assigned $j = 1$, is a symmetric mode. Moreover, for an $r$-fold degenerate eigenvalue of a particular
mode (degeneracies may also arise due to number-theoretic origins), it is arbitrarily chosen to define $j$ by ordering the $r$ degenerate eigenvalues in an increasing order of the quantum number $n$. This
convention effectively orders the degeneracies in an increasing order of $m$ as well so the two quantum numbers are essentially equivalent with regard to the choice. It was shown by Courant that for a
domain on $\mathbb{R}^2, \nu_j \le j$ $\forall j \in \mathbb{N}$ \cite{courant} and the arrangement of the spectrum respects this statement.
\\
\\
\noindent
The pattern of nodal domains for the anti-symmetric mode may be constructed through the juxtaposition of the nodal domain patterns of two triangles with angles $30^{\circ}\--60^{\circ}\--90^{\circ}$ that
are exactly antisymmetric with respect to each other as illustrated by Figure \ref{examples}. Hence, we consider only the eigenfunctions of the symmetric mode of vibration, defined above by
$\psi_{m,n}^c$, hereafter referred to as simply $\psi_{m,n}$.

\section{The pattern of nodal domains}
\begin{figure} [H]\scriptsize
\begin{center}
\subfloat[]{\scalebox{0.3}{\includegraphics{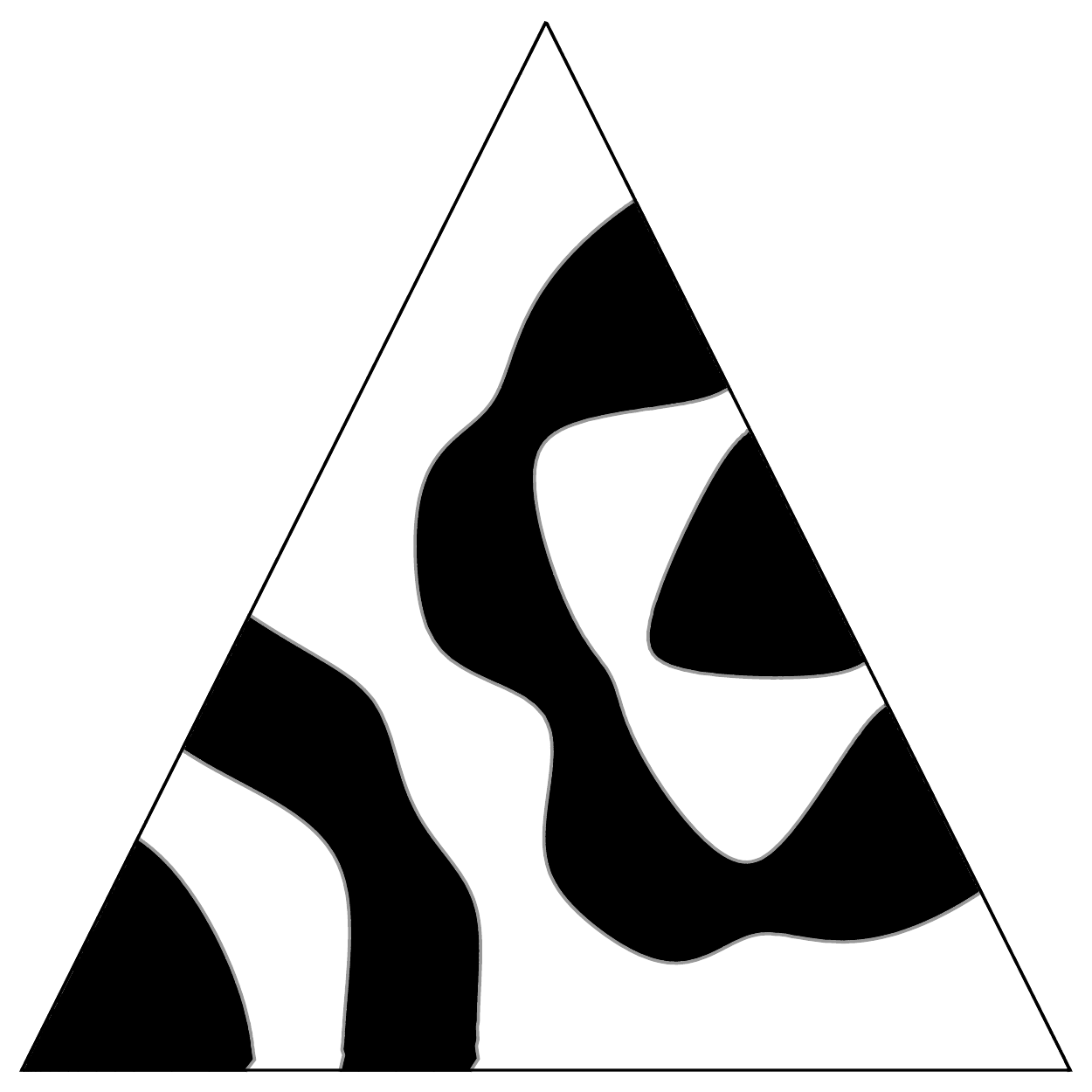}}}
\qquad \qquad
\subfloat[]{\scalebox{0.3}{\includegraphics{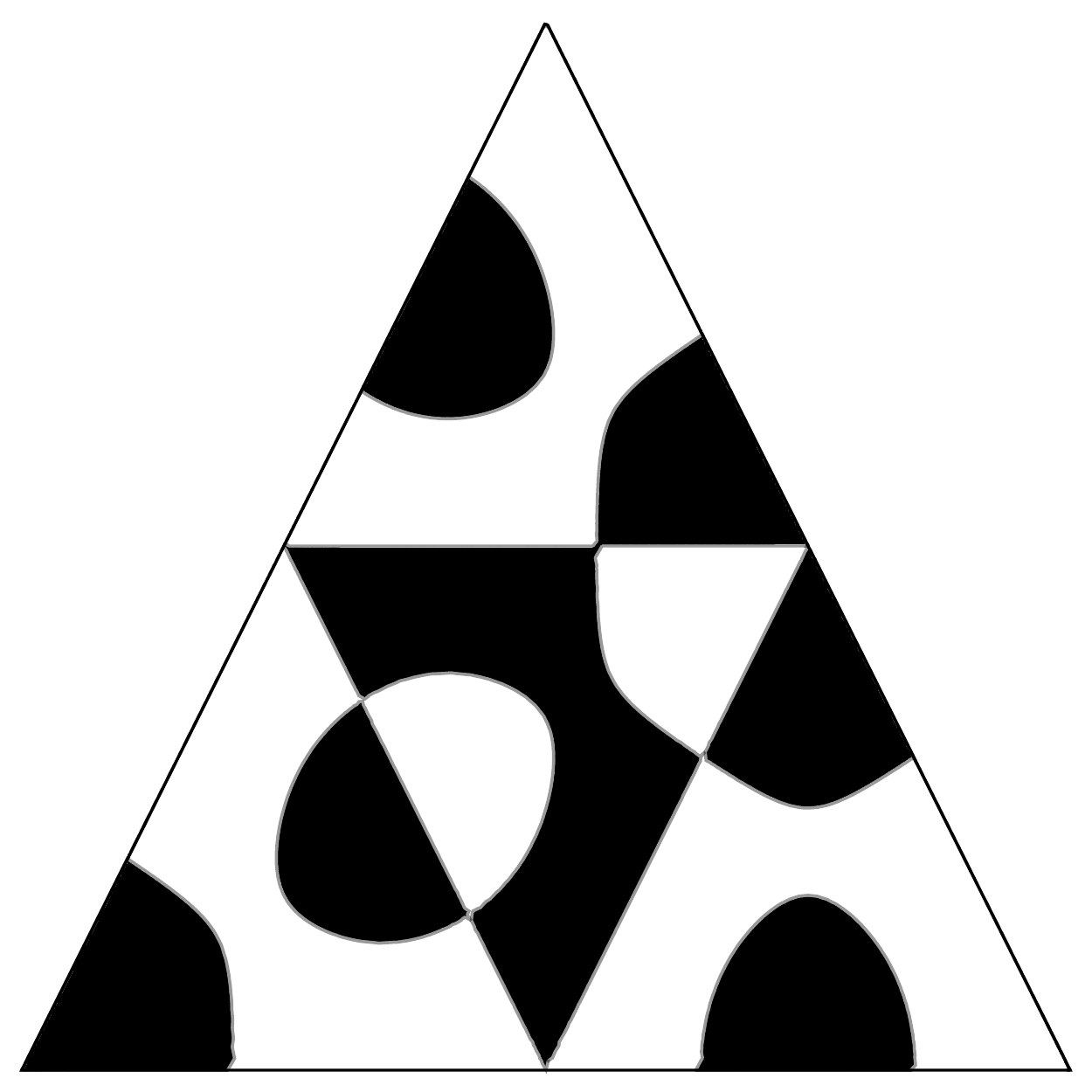}}}
\qquad \qquad
\subfloat[]{\scalebox{0.3}{\includegraphics{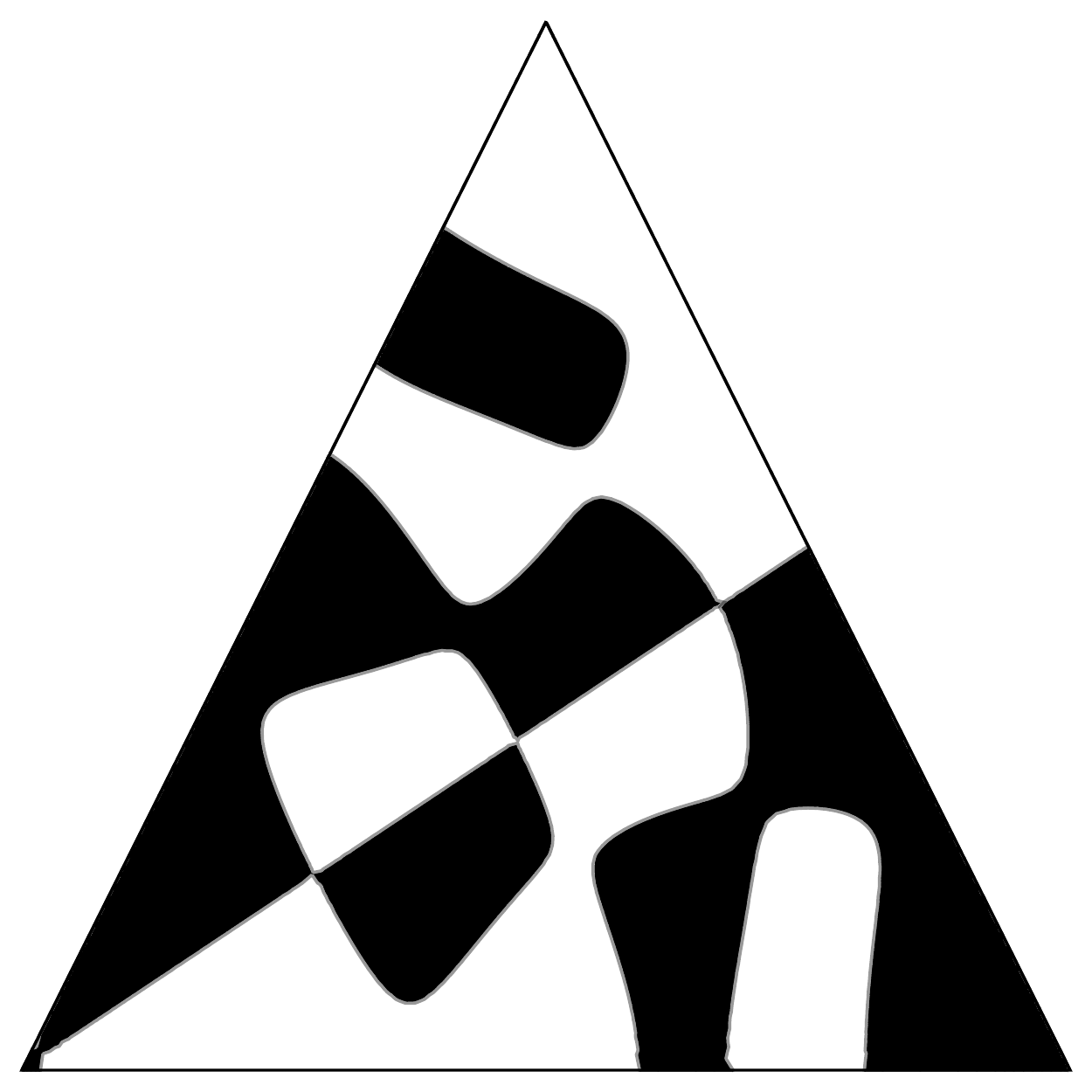}}}
\end{center}
\caption{\footnotesize (a) The pattern of nodal domains for $\psi_{8,3}^c$. (b) The tiling structure is explicitly shown in the pattern of $\psi_{8,2}^c$, which is constructed by the juxtaposition of four tiles. (c) The antisymmetric mode of vibration corresponding to the wavefunction $\psi_{8,3}^s$. The wavefunction is positive in the white areas and negative in the darkened regions. }
\label{examples}
\end{figure}

\noindent
The nodal pattern of the eigenfunctions of the equilateral triangle exhibits certain symmetry relations, one of which is reflected in a tiling structure of the nodal lines. The nodal set of the eigenfunction $\psi_{m,n}$ with $m \ge 2n$ and $gcd (m,n) = d > 1$ is composed of $d^2$ identical nodal patterns which are similar to the nodal set of the eigenfunction $\psi_{m',n'}$ with $m' = m/d$ and $n' = n/d$. This arrangement, which directly follows from \eqref{eq: wf}, is illustrated by the examples in Figure \ref{examples}. \\
\\
Each of the tiles corresponding to the nodal pattern of $\psi_{m',n'}$ is contained within an equilateral subtriangle of area $\displaystyle \frac{\sqrt{3}\pi^2}{4 \,d^2}$. The complete nodal pattern is constituted by these subtriangles which together tessellate the plane of the equilateral triangle upon reflection. The tiling nature of the wavefunctions when $gcd (m,n) \ne 1$ is by no means an exclusive feature of the equilateral triangle billiard only \-- the same behaviour has also been demonstrated for the isosceles right-angled triangle by \cite{aronovitch}.
\noindent
Since the number of nodal domains $\nu_{m,n}$ may be easily obtained from the nodal domain count for the reduced values $m' \mbox{ and } n'$ through the relation $\nu_{m,n} = d^2\,\nu_{m',n'}$, it suffices to consider the nodal count for the non-tiling case wherein $d = gcd (m,n) = 1$. The following Sections maintain this implicit assumption and we discuss only non-tiling wavefunctions,
unless  mentioned otherwise. It is often simpler to decompose the nodal count in terms of the number of boundary intersections ($\tilde{\nu}_{m,n}$) and nodal loops ($I_{m,n}$) in accordance
with the approach outlined in \cite{aronovitch}. $\tilde{\nu}_{m,n}$ represents the number of intersections of the nodal set with the boundary $\partial {\cal D}$ of the 2-D manifold i.e. the number of times where the $boundary function$, the normal derivative of the wavefunction under Dirichlet conditions, vanishes on $\partial {\cal D}$. Nodal loops may be recognised as nodal curves  (``lines") which neither touch the boundary nor intersect themselves or any other nodal line. The nodal loops therefore enclose interior domains whose boundaries do not include the boundaries of the equilateral triangle itself. For a non-tiling wavefunction $\psi_{m',n'}$, the relation between these terms is \cite{aronovitch}
\begin{equation}
\label{eq: total}
\nu_{m,n} = 1 + \frac{1}{2}\tilde{\nu}_{m,n} + I_{m,n}.
\end{equation} 
The nodal domain count can thus be obtained by combining the boundary intersections count and the nodal loop count in \eqref{eq: total}. We present certain partial formulae for the total number of nodal
domains and the nodal loop count along with an exact formula for $\tilde{\nu}_{m,n}$  in the following subsections.

\subsection{The number of nodal domains}
\noindent 
The nodal domain count may be numerically determined for a given function using the Hoshen-Kopelman algorithm \cite{hk}. The applicability of the algorithm stems from the convenient fact that the nodal
lines of the equilateral triangle do not intersect under non-tiling circumstances. Although the eigenfunction is represented on a grid of a finite resolution, accuracy of the count is maintained by
ensuring that the resolution is sufficiently large so as to distinguish between nodal lines near avoided intersections. An instructive critique may be seen in \cite{monastra}.\\
\\Each non-tiling wavefunction $\psi_{m,n}$, for a fixed value of $n$, may be categorised into one of at most $3(n-1)$ equivalence classes depending on the value of the residue $c = m\mod3n$. The
sequence of nodal domain counts for each class, when analysed with regard to the second difference of $\nu_{m,n}$, shows the existence of the recurrence relation
\begin{equation}
\label {eq: recurrence}
\nu_{m+6n,n} - 2 \nu_{m+3n,n} +\nu_{m,n} = 3n^2.
\end{equation}

{\small
\begin{center}
\begin{longtable}{c c c | c c c}
\hline
$m$ & $n$ & $c = m\mod 3n$ & $\nu_{m,n}$ &  $\Delta\:\nu_{m,n} $ & Second difference \endhead \hline
7 & 2 & 1 & 6 & \--- & \---\\
13 & 2 & 1 & 21 & 15 & \---\\
19 & 2 & 1 & 48 & 27 & 12\\
25 & 2 & 1 & 87 & 39 & 12\\
31 & 2 & 1 & 138 & 51 & 12\\
37 & 2 & 1 & 201 & 63 & 12\\
9 & 2 & 3 & 10 & \--- & \---\\
15 & 2 & 3 & 29 & 19 & \---\\
21 & 2 & 3 & 60 & 31 & 12\\
27 & 2 & 3 & 103 & 43 & 12\\
33 & 2 & 3 & 158 & 55 & 12\\
39 & 2 & 3 & 225 & 67 & 12\\
\hline
\caption{An illustration of the constancy of the second difference of the total number of nodal domains when the wavefunctions belonging to the same class, defined by $m\mod3n$, are arranged in increasing order of the quantum number $m$. The difference between succesive values of $m$ considered in any such sequence is naturally $3n$. It is to be noted that the value of the second difference is independent of the class and is given by $3n^2$.}
\end{longtable}
\end{center}
}
\setlength{\tabcolsep}{31.5pt}
\renewcommand{\arraystretch}{2.5}
This observation stems from an extensive investigation over 8050 eigenfunctions. The equation \eqref{eq: recurrence} can be analytically solved to yield the following result:
\begin{equation}
\label{eq: diff_sol}
\nu_{m,n} = \frac{3}{2} \bigg(\frac{m^2}{9} - \frac{mn}{3}\bigg) + \frac{m\alpha}{3n} + \beta,
\end{equation}
where $\alpha$ and $\beta$ are two parameters dependent on $c$ and $n$. Comparison of \eqref{eq: diff_sol} with the tables of evaluated domain counts clearly indicates $\displaystyle \alpha = \frac {(3n -
n^2)}{2}$. This, when coupled with further observations about the nature of the parameter $\beta$, helps to effectively reduce \eqref{eq: diff_sol} into two major cases:
\begin{eqnarray}
\label{eq: approx}
\nu_{m,n} &=& \frac{m^2}{6} - \frac{(4n-3)m}{6} + n^2 - \frac{cn - F_1(c, n)}{3} \hspace{1.5cm} \mbox{if} \hspace{0.3cm} 0 < c < n,
\nonumber \\&=& \frac{m^2}{6} - \frac{(4n-3)m}{6} + n^2 - \frac{2(c-n)n - F_2(c, n)}{3} \hspace{0.3cm} \mbox{if} \hspace{0.3cm} n < c < 3n.
\end{eqnarray}
Although the explicit forms of the functions $F_1$ and $F_2$ are not exactly known, certain functional relations and properties are observed to be satisfied, which are as enlisted below:
\begin{enumerate}
\item $F_1(c, \, c+1) \:\:= \,c^2 + 3$.
\item $F_1(c, \,2c+1) = F_1(c,\, 2c+2) = c(c+3)$.
\item
$F_1(c, n) =
\begin{cases}
F_1(c,\, 7c - n) & \hspace{1cm}\text{for }\hspace{1cm}2c < n < 3c, \\
F_1(c,\, 3c+ n) & \hspace{1cm} \text{for } \hspace{1cm}2c < n.
\end{cases}$
\end{enumerate}
Therefore, an exact formula for the number of domains is obtained whenever $n$ equals $c+1, \,2c+1 \mbox{ or } 2c+2$. The existence of an exact formula has also been verified for all cases whenever $c = 1\: (0<c<n)$, irrespective of the value of $n$, and is given by
\begin{equation}
\label{eq: exact}
\nu_{m,n} = \frac{m^2}{6} - \frac{(4n-3)m}{6} + n^2 - \frac{n - 4}{3}.
\end{equation}
\begin{figure} [H]
\begin{center}
\scalebox{0.76}{\includegraphics{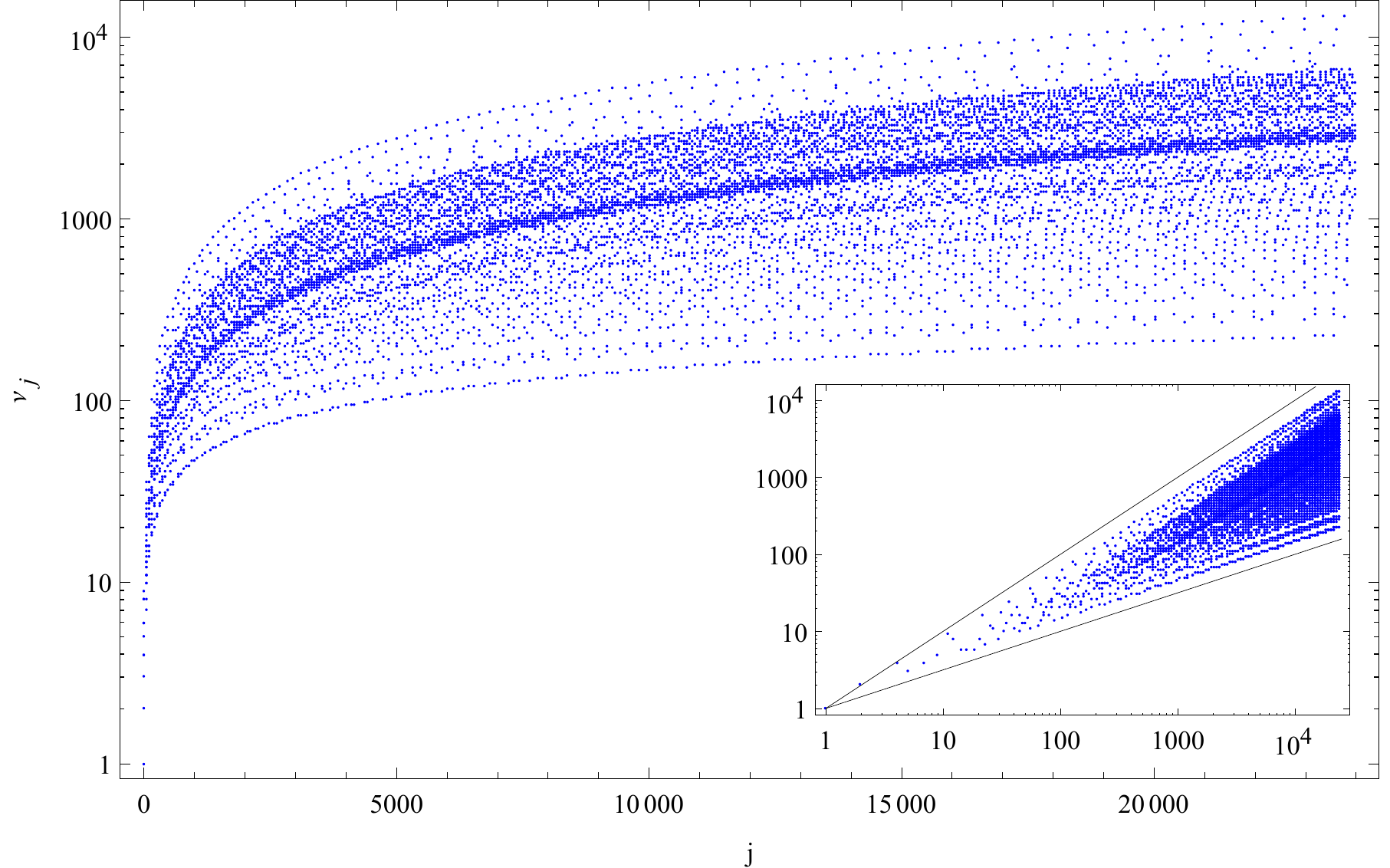}}
\end{center}
\caption{The number of nodal domains for the equilateral triangle billiards for the first 12045 wavefunctions (in increasing order of energy). Inset: The corresponding plot of log $\nu_j$ against log $j$. Both the figures clearly indicate a lower and upper bound for the values of $\nu_j$ as a function of $j$. The inset figure, bounded by straight lines of slopes 0.5 and 1, shows the scaling of $\nu_j$ with $j$ as $j\rightarrow \infty$. }
\label {Nuj}
\end{figure} 
Moreover, a reasonable approximation to the total number of domains can be made by substituting $F_1(c, \,2c+1)$ in lieu of $F_1(c, n)$ in \eqref{eq: approx} since the value thus obtained is always an upper bound to $\nu_j$ for any given values of $c$ and $n$ satisfying $0 < c < n$. Similar investigations into the nature of $F_2$ show that $F_1(c_1, n)$ and $F_2(c_2 - n,\, n)$ (where $0 < c_1 < n$ and  $n < c_2 < 3n$) share common elements in their sets of possible values for a given class, but are not identical, if $c_1 = c_2 - n$.\\ 
\\The number of nodal domains for the equilateral triangle billiard is shown in Figure \ref{Nuj} with reference to the variation of $j$ for the first 12045 wavefunctions of the symmetric mode. The corresponding plot of log $\nu_j$ against log $j$ is clearly bounded by two straight lines of slopes nearly equal to $0.5$ and $1$. This proves the simultaneous presence of both the scalings $\displaystyle \nu_{j} \sim j^{1/2}$ and $\displaystyle \nu_{j} \sim j$ for high $j$, which hold for boundary domains and interior domains respectively.

\subsection{The boundary intersection count}

\noindent 
The number of boundary intersections of the wavefunction encodes the characteristics of the classical ray dynamics of the billiards system since the wavefunction is completely defined by its
corresponding boundary function. Extensive analysis of numerical data reveals that in the absence of a tiling structure,
\begin{equation}
\label{eq: bdry_spl}
\tilde{\nu}_{m,n} = \,2m - 4 - 4\delta_{\{(m+n)\mod3\}^2 + \{(m-n+1)\mod3\}^2, 0},
\end{equation}
where $\delta_{i,j}$ denotes the Kronecker symbol. However, in the tiling case i.e. when $d \ne 1$,
\begin{equation}
\label{eq: bdry_main}
\tilde{\nu}_{m,n} = \tilde{\nu}_{m',n'} + (d-1)\bigg\{(2m'-1) - 4\delta_{\{(m'+n')\mod3\}^2 + \{(m'-n'+1)\mod3\}^2, 0}\bigg\}.
\end{equation}
\\
It is apparent that \eqref{eq: bdry_main} simply reduces to \eqref{eq: bdry_spl} for the non-tiling case when $d = 1$. Therefore, $\tilde{\nu}_{m,n}$ for the eigenfunctions having a tiling structure may
be represented as a sum of $\tilde{\nu}_{m',n'}$ and a certain number of multiples of a definite step size.
\begin{figure} [H]
\begin{center}
\subfloat[]{\scalebox{0.65}{\includegraphics{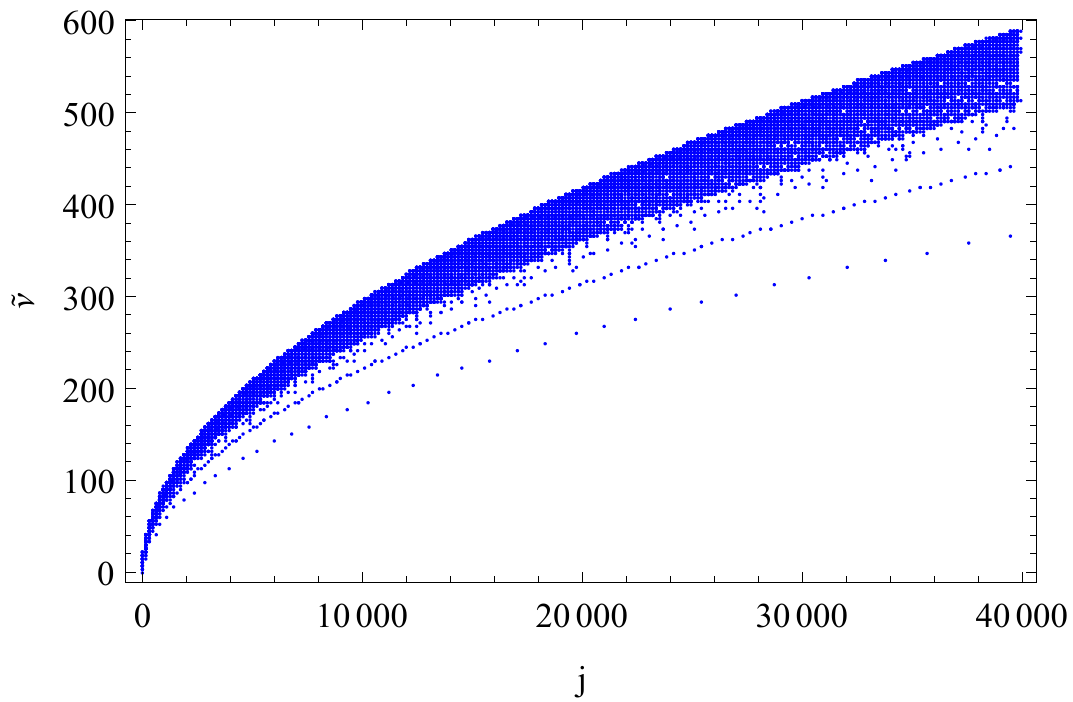}}}
\subfloat[]{\scalebox{0.66}{\includegraphics{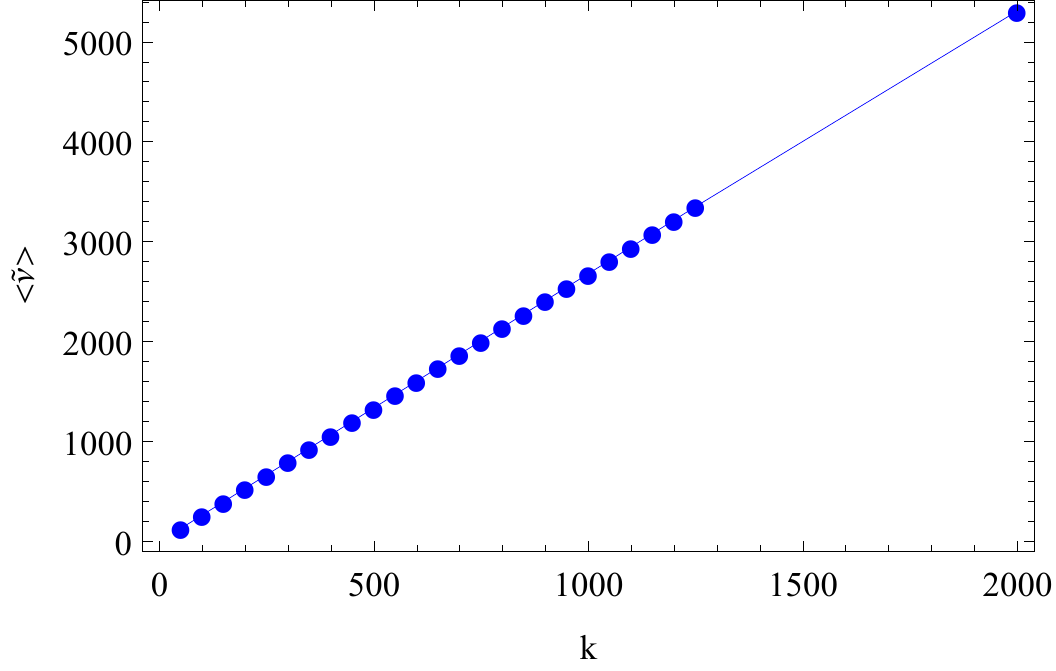}}}
\end{center}
\caption{(a) A plot of the number of boundary intersections $\tilde{\nu_j }$ for the lowest 30500 eigenfunctions of the equilateral triangle billiard, showing the scaling $\tilde{\nu_j } \sim j^{1/2}$ as expected from observations of other billiard systems. (b) The linear dependence of $\langle\tilde{\nu}\rangle$, the number of intersections averaged over the spectral interval, with $k = \sqrt{E}$ as shown verifies this proposed scaling since $j$ is determined by the increasing order of $E$ or specifically, $E \sim j$.}
\label{Tilde}
\end{figure} 
\noindent
Let $I_{\lambda}(E)$ be the spectral interval $ [E, E+\lambda E] \mbox{ where } \lambda > 0$ is arbitrary. $N_{I}$ denotes the number of eigenvalues of the wavefunction in the interval $I_{\lambda}(E)$.
It has been shown in \cite{toth} that the number of boundary domains $\tilde{\nu}_{j}$, which is equal to the number of boundary intersection points in two dimensions, is $O(\sqrt{j})$. The variation of
$\tilde{\nu}_{j}$ with $j$ for the triangle billiards is plotted in Figure \ref{Tilde}(a) and the corresponding representation of both quantities on a logarithmic scale confirms the aforementioned scaling.
Furthermore, averaged over the spectral interval, $\langle\tilde{\nu}\rangle$ is studied as a function of $k = \sqrt{E}$ and the results thereof are presented in Figure \ref{Tilde}(b). We observe that the graph is very nearly perfectly linear, thereby
indicating that $\langle\tilde{\nu}\rangle \propto k$ akin to the predictions of the random wave model for chaotic billiards where $\displaystyle \langle\tilde{\nu}\rangle \approx k
\emph{L}/{2\pi}$ as proved by \cite{uzy2002}. This proportionality also establishes that $\tilde{\nu}_{j} \sim O(\sqrt{j})$. However, the dissimilarity with the model arises from the fact that
$var(\tilde{\nu})$ is not directly proportional to $k$ for the equilateral triangle. We return to the statistical properties in the next Section. 

\subsection{The nodal loop count}
The nodal loop count for a wavefunction $\psi_{m,n}$ is easily evaluated from the total number of nodal domains and the number of boundary intersections, which is computed by \eqref{eq: bdry_main}. It is
important to note that the loop count is zero if and only if $\displaystyle n = \bigg\lfloor\frac{m-1}{2}\bigg\rfloor$. As in the case of the total number of domains, let the wavefunctions be categorised
into equivalence classes based on the value of $c = m\mod 3n$. A similar analysis of the second difference of the loop counts reveals that the same recurrence relation is satisfied by the loops as
well. Therefore,
\begin{equation}
I_{m+6n,n} - 2 I_{m+3n,n} +I_{m,n} = 3n^2.
\end{equation}
This equation has the same general solution as \eqref{eq: recurrence}. However, it is much more insightful to consider the sequence $\{I_{m,n},\, I_{m+\gamma n, n},\, I_{m+2\gamma n, n},\,\dots\}, \gamma \in \mathbb{N}$ and $ m/\gamma \in \mathbb{N}$. The arrangement also ensures that all the wavefunctions corresponding to the loop counts in any sequence belong to the same residue class $c_{I} = m\mod n$. It is further observed that
\begin{eqnarray}
\label{eq: gamma}
I_{m+2\gamma n,n} - 2 I_{m+\gamma n, n} + I_{m,n} &=& \frac{\gamma^2 \,n^2}{3},   \hspace{1.3cm} \mbox{if} \hspace{1cm}\gamma =3 z,\, z \in \mathbb{N};
\nonumber \\&=& \theta_{i}, \: \hspace{0.4cm}1\le i\le3, \hspace{0.9cm} \mbox{otherwise.}
\end{eqnarray}
The set $\theta = \{\theta_{1}, \, \theta_{2}, \, \theta_{3}\}$ is naturally a characteristic of $n$, $\gamma$ and $c_{I}$ and possesses the property that $\textstyle \sum_{i = 1}^3 \theta_{i} =
\gamma^2 n^2$. The transformation $\displaystyle \vartheta_{i} = \theta_{i} - \frac{\gamma^2 \,n^2}{3}$ maps the set $\theta$ to the set $\vartheta$, which bears the obvious property that all of the elements
$\vartheta_{i}$, being the deviations from a certain average, sum to zero. This implies,
\begin{equation}
I_{m+2\gamma n,n} - 2 I_{m+\gamma n, n} + I_{m,n} = \vartheta_{i} +  \frac{\gamma^2 \,n^2}{3}, \hspace{1cm} \mbox{if} \hspace{1cm}\gamma \ne 3 z,\, z \in \mathbb{N}.
\label{eq: omega}
\end{equation}
Additionally, $\vartheta$ is independent of $\gamma$ but is identical for $c_{I}$ and $n - c_{I}$, provided that $n$ remains
constant. The second difference is also periodic with a period of $3\gamma n$ such that $\theta_{i}$ as well as $\vartheta_{i}$ is the same for $ I_{m,n} \mbox{ and } I_{m + 3\gamma n,n}$. We show in the following Section that the set $\vartheta$ encodes the point group symmetry of the equilateral triangle that is visible in the pattern of the nodal domains.

\subsection{Origin of $C_{3}$ point group symmetry}
\begin{figure} [H]
\begin{center}
\subfloat[]{\scalebox{0.3}{\includegraphics{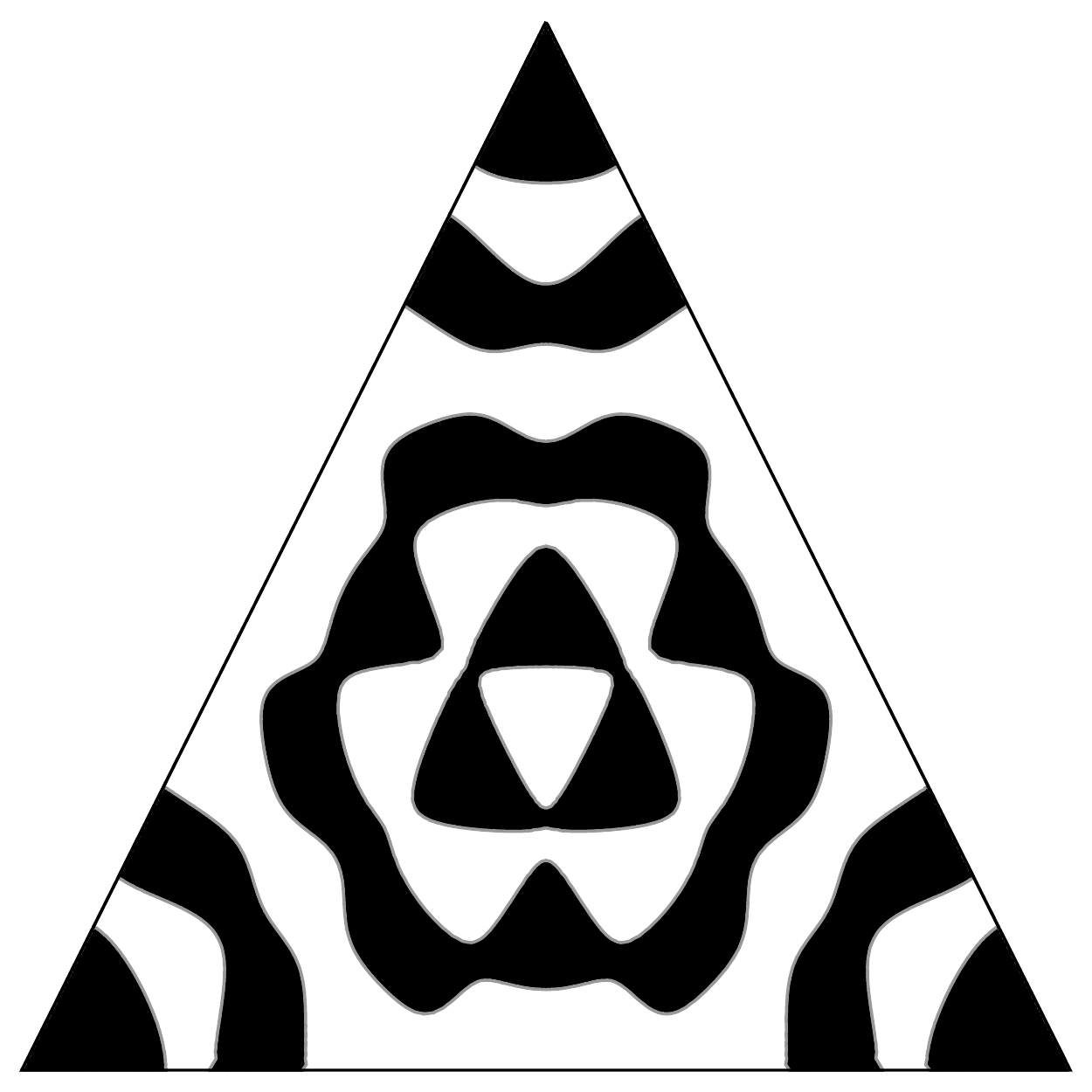}}}
\qquad \qquad \qquad
\subfloat[]{\scalebox{0.3}{\includegraphics{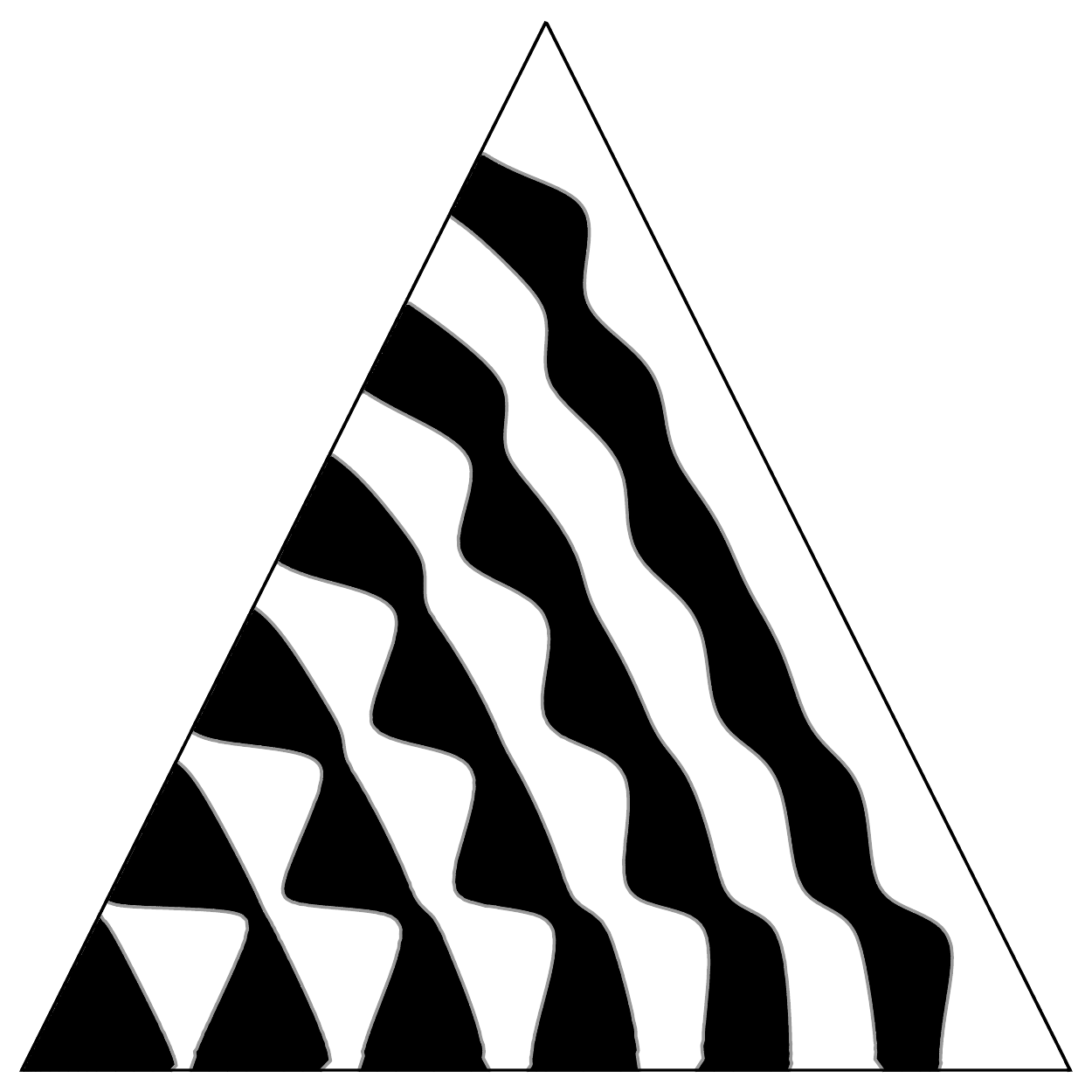}}}
\qquad \qquad
\end{center}
\caption{(a) A plot of the nodal structure of the wavefunction $\psi_{13,5}$, which belongs to the A1 class. The functions of this class are necessarily symmetric under the operation $R_{120}$ of clockwise rotation through $120^\circ$ about the incentre of the triangle . (b) The pattern of domains for $\psi_{13,6}$, which exhibits the properties of the $E$ symmetry class. The function is symmetric about $R_{1}$ and under no other operation, excluding $I$. The operations $R_{120}$ and $R_{1}$ are as defined below. All wavefunctions are positive in the white areas and negative in the darkened regions. }
\label{A1E}
\end{figure} 
All the eigenfunctions of the symmetric mode of the equilateral triangle billiards may be classified into the $A1$ or $E$ classes of the $C_{3v}$ point group \cite{canada} based on the symmetry properties of their nodal structure.
Let the vertices of an equilateral triangle be labelled with the three possible values of $m_{i}\mod3$, where $m_{i} \in \{m,\, m+3\gamma n,\, m+ 6\gamma n,\, \dots \}$ and $ m/\gamma \in \mathbb{N}$. Let $R_{i}\, (i = 0,\,1,\,2)$ connote reflection of the triangle about the line passing through the vertex $i$ and perpendicular to the opposite side. The operation of rotation of the triangle through an angle $z^{\circ}$ clockwise about the incentre is denoted by $R_{z} \,(z = 120,\,240)$. These five symmetry operations together with the identity operation $I$ constitute the $C_{3v}$ group. The vertex marked with $m_{i}\mod3$ is assigned the value of $\vartheta_{i}$, where $m_{i} \mbox{ and } \vartheta_{i}$ correspond to $m \mbox{ and } \vartheta_{i}$ in \eqref{eq: omega} respectively. If the notation $(\vartheta_{1},\,\vartheta_{2},\,\vartheta_{3})$ is used to indicate that $\vartheta_{1},\,\vartheta_{2}\, \mbox{ and } \vartheta_{3}$ are assigned to the vertices with labels 0, 1 and 2 respectively, then these six operations are mathematically defined as follows:
\\
\begin{eqnarray*}
I &:& \hspace{1cm} (\vartheta_{1},\,\vartheta_{2},\,\vartheta_{3}) \: \rightarrow \: (\vartheta_{1},\,\vartheta_{2},\,\vartheta_{3})
\\R_{120} &:&\hspace{1cm} (\vartheta_{1},\,\vartheta_{2},\,\vartheta_{3}) \: \rightarrow \: (\vartheta_{2},\,\vartheta_{3},\,\vartheta_{1})
\\R_{240} &:&\hspace{1cm} (\vartheta_{1},\,\vartheta_{2},\,\vartheta_{3}) \: \rightarrow \: (\vartheta_{3},\,\vartheta_{1},\,\vartheta_{2})
\\R_{0} &:&\hspace{1cm} (\vartheta_{1},\,\vartheta_{2},\,\vartheta_{3}) \: \rightarrow \: (\vartheta_{1},\,\vartheta_{3},\,\vartheta_{2})
\\R_{1} &:&\hspace{1cm} (\vartheta_{1},\,\vartheta_{2},\,\vartheta_{3}) \: \rightarrow \: (\vartheta_{3},\,\vartheta_{2},\,\vartheta_{1})
\\R_{2} &:&\hspace{1cm} (\vartheta_{1},\,\vartheta_{2},\,\vartheta_{3}) \: \rightarrow \: (\vartheta_{2},\,\vartheta_{1},\,\vartheta_{3})
\end{eqnarray*}
\\
As $\gamma$ is changed, keeping in consideration the enforcement of non-tiling conditions, the three values of $\vartheta_{i}$ are permuted with respect to the vertices labelled by $m_{i}\mod3$. Any such transformation brought about by the variation in $\gamma$ is exactly equivalent to one of the three rotation operations that constitute the $C_3$ group, a simple subgroup of the $C_{3v}$ group. Most importantly, all such transformations obey the composition law of the group. Let $\gamma = i$ and $\gamma = f$ be the initial and final states of the system, denoted as $\gamma _i$ and $\gamma _f$ respectively. Then the series of $N$ operations 
\begin{equation*}
\gamma_{i}\: \xrightarrow[]{O_{1}} \: \gamma_{z_{1}} \: \xrightarrow[]{O_{2}} \: \gamma_{z_{2}} \xrightarrow[] {O_{3}} \: \dots  \: \xrightarrow[]{O_{N-2}} \: \gamma_{z_{N-2}} \: \xrightarrow[]{O_{N-1}} \: \gamma_{z_{N-1}} \: \xrightarrow[]{O_{N}} \: \gamma_{f}, 
\end{equation*}
where $\gamma_{z_{i}}$ are the intermediate states, is equivalent to a single symmetry operation of the $C_{3}$ group defined as
\begin{equation*}
O := O_{N}\:\circ O_{N-1} \:\circ O_{N-2} \:\circ \: \dots \:O_{3} \:\circ O_{2} \:\circ O_{1}.
\end{equation*}

\begin{figure} [H]
\begin{center}
\scalebox{0.55}{\includegraphics{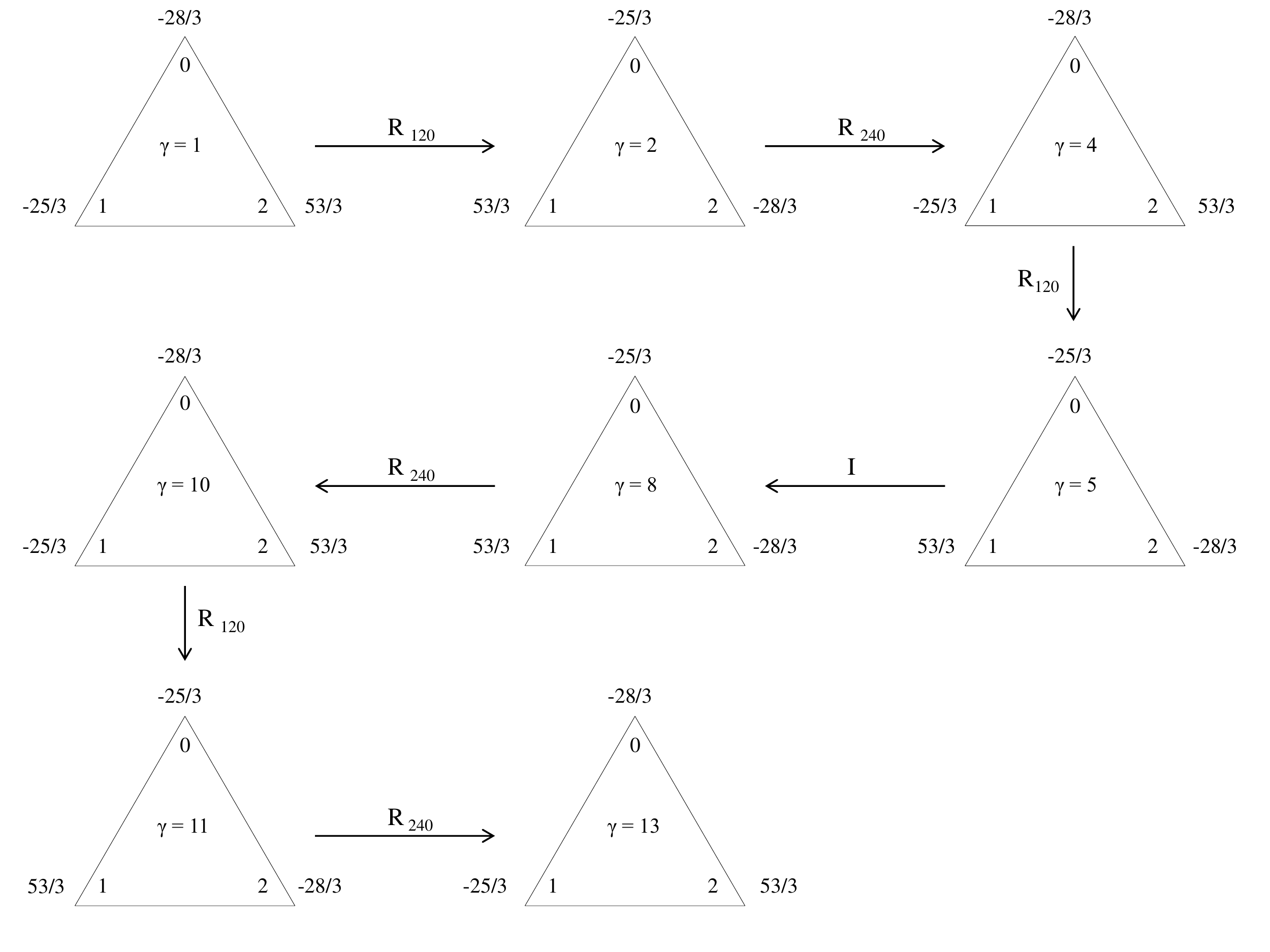}}
\end{center}
\caption{The permutation of the labels $\vartheta_{i}$, with respect to the vertices labelled by $m_{i}\mod3$, caused by the variation in $\gamma$ for $n = 7$ and $c_I = 1$. The operation effecting each transformation is indicated. The transformation $\gamma \rightarrow \gamma +1 $ is always performed by the $R_{120}$ operation in the example shown. It is easy to verify that the transformations obey the composition law of the symmetry operations of the $C_{3}$ group.}
\label {symm}
\end{figure} 
The concept is illustrated by Figure \ref{symm} for $n = 7$ and $c_I = 1$. For instance, the transformation from $\gamma = 4 \mbox{ to } \gamma = 10$ may either be effected throught the series of operations shown, or through the single identity operation $I$, which is equivalent to the composition of the three. Similarly, the transformation from $\gamma = 1 \mbox{ to } \gamma = 11$ is enacted by a single application of $R_{120}$ which is obtained mathematically as the composition of the operations outlined in the diagram. Only those values of $\gamma$ for which $\gamma \ne 3z, z\in \mathbb{N}$ are represented in the diagram in accordance with \eqref{eq: gamma}.\\
\\
Figure \ref{symm} may be extended to higher values of $\gamma$ but the fingerprints of the $C_{3}$ symmetry continue to persist. Considering $z \in \mathbb{N}$, let the \emph{ fundamental operation} for a particular $n$ be defined as $R_{120}$ if $n = 3z +1$ and $R_{240}$ if $n = 3z -1.$ The case when $n = 3z$ is deliberately not considered as all the members of any class $c_I$ have the same value of $m\mod3$ in such a situation. The fundamental operation, denoted by $O_{F}$, effects the transformation $\gamma \rightarrow \gamma + 1$. Thereafter, the transformation $\gamma \rightarrow \gamma +z,\, z\in \mathbb{N}$ is effected by the single operation $(O_{F})^z$ which is equivalent to the composition of $z$ succesive applications of $O_{F}$. Owing to this property, the transformation $\gamma \rightarrow \gamma +3$ is always defined by the identity operation.\\
\\Two further examples of these group-theoretic considerations are presented in the Appendix A for further elucidation and corroboration of the idea. We believe that this observation is quite beautiful, and reassuring in the group-theoretic sense.

\section{Nodal domain statistics}
The statistics of nodal domains is extremely important for the analysis of the equilateral triangle billiard. Statistical analysis of the eigenfunctions of equilateral triangle billiard clearly establishes its distinctiveness with reference to both chaotic and separable systems. At the same time, it also presents an image of the underlying classical ray dynamics of the triangle. We present the results obtained from the  numerical experiments. Along with these, we also present some analytical results for the statistics on the count of boundary intersections. 

\subsection{The normalised number of boundary intersections}

The normalised number of nodal intersections with the boundary associated with each wavefunction is defined as $\displaystyle {\eta_{j} = \frac{\tilde{\nu}_{j}}{\sqrt{j}}}$, since the appropriate parameter for normalisation was verified to be $\sqrt{j}$  above. The $\eta$ distribution for $I_{\lambda}(E)$, which is a characteristic of the system, is
\begin{equation}
P[\eta, I_{\lambda}(E)] = \frac{1}{N_{I}} \sum_{E_{j} \in I_{\lambda}(E)} \delta \bigg(\eta - \frac{\tilde{\nu}_{j}}{\sqrt{j}} \bigg) = \frac{1}{N_{I}} \sum_{E_{j} \in I_{\lambda}(E)} \delta (\eta -
\eta_{j}).
\label{eq: eta}
\end{equation}
Hence, the limiting distribution of $P$ for the system is 
\begin{equation}
P(\eta) = \lim_{E \rightarrow \infty} P[\eta, I_{\lambda}(E)].
\end{equation}
\\Unlike the distribution for chaotic billiards that limits to a single-peak Dirac delta function centered at $\displaystyle \eta = \frac{\emph{L}}{\sqrt{\pi {\cal A}}}$ (where \emph{L} is the perimeter of the billiard) or separable systems which exhibit monotonic behaviour, this distribution for 882455 eigenfunctions is composed of multiple peaks of different strengths at certain characteristic values of $\eta $. Since the length of the boundary $L$ serves to only scale the wavefunction over the domain ${\cal D}$, as may be seen from \eqref{eq: wf}, but does not affect the fundamental pattern of the nodal domains per se, the boundary intersections count is independent of the length of the sides. We therefore suggest that the three distinctly sharp peaks of $P(\eta)$ correspond to certain geometric quantities related to the equilateral triangular domains.
\begin{figure} [H]
\begin{raggedleft}
\scalebox{0.567}{\includegraphics{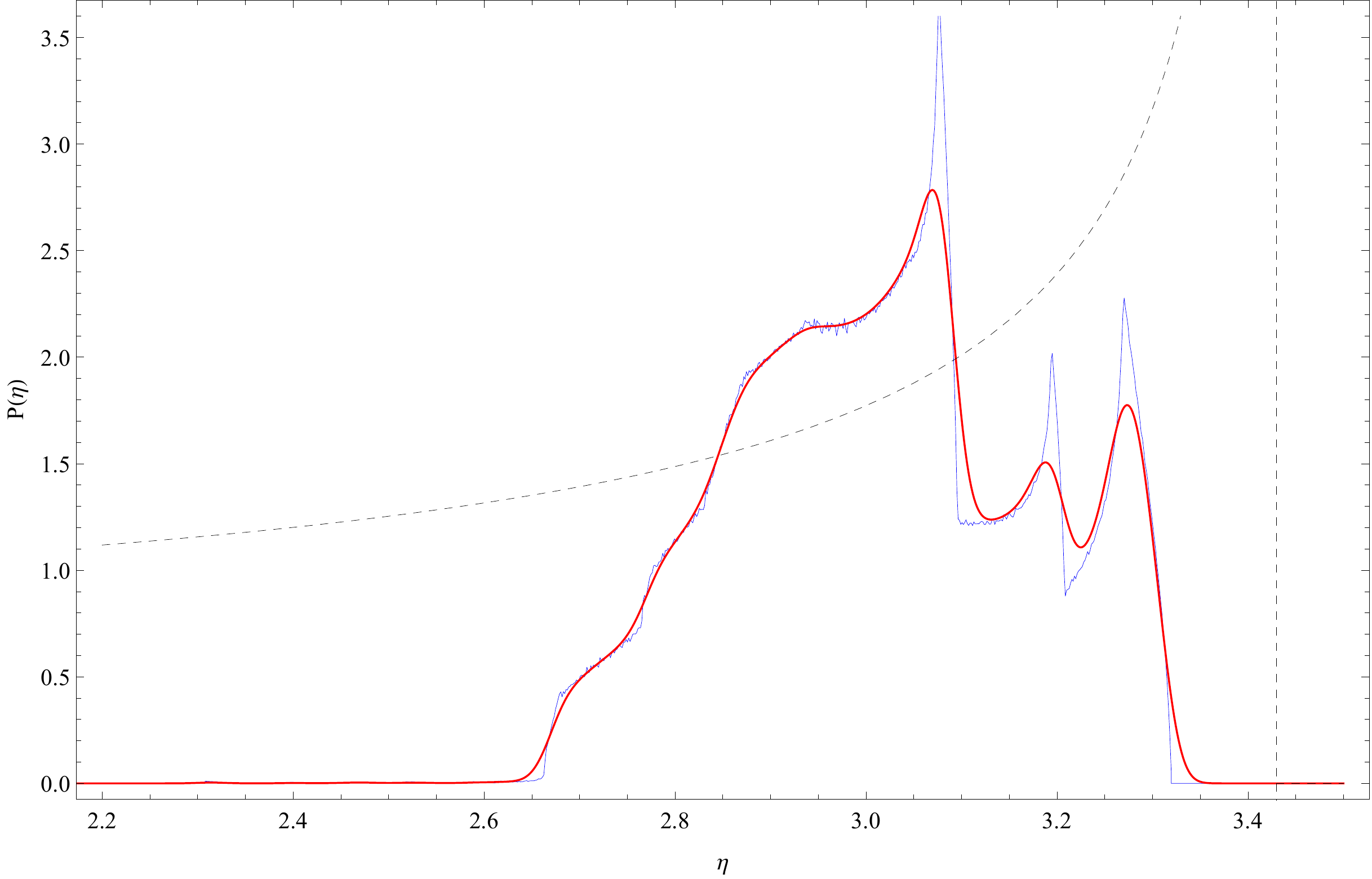}}
\end{raggedleft}
\caption{The distribution, $P(\eta)$, of the normalised number of boundary intersections for the system shown for $\lambda = 1$ and $E = 2000^{2}$, considering 882455 wavefunctions in the spectral interval. The blue curve shows the exact distribution as a function of $\eta$ whereas the red curve (in bold) depicts the smooth histogram representation of the same and serves to clearly distinguish the most dominant peaks. The dotted line corresponds to the distribution predicted by the analytical result of the following section. }
\label {PEta}
\end{figure} 
\subsubsection{Analytical treatment}

We follow the methods outlined by \cite{shankar,karageorge} in this development. As discussed in \cite{shankar}, we begin by the parametrization of the spectrum by introducing  the transformations 
\begin{equation}
m = \sqrt{E}\cos{\theta} \mbox{ and } n = \displaystyle {\frac{\sqrt{E}\big(\cos{\theta} + \sqrt{3 \,{\sin^2{\theta}+1}}\,\big)}{2}}. 
\end{equation}
It is observed that $m^2+n^2 - mn = E$. The cumulative density of energy levels follows the Weyl formula:
\begin{equation}
N(E) \approx \displaystyle{\frac{AE-L\sqrt{E}}{4\pi}} =  \displaystyle{\frac{\sqrt{3}\pi E}{16}} \bigg(1- \displaystyle{\frac{4\sqrt{3}}{\pi\sqrt{E}}}\bigg).
\end{equation}
Consider only the non-tiling case, we have $\tilde{\nu}_j = 2m -4 \mbox{ or } 2m-8$ as is appropriate. Therefore, 
\begin{equation}
\tilde{\nu}_j = 2\sqrt{E}\cos{\theta}-4 \mbox{ and } \eta_j(E,\theta) = \displaystyle{\frac{\tilde{\nu}_j}{\sqrt{j}}}= \displaystyle{\frac{8(\cos{\theta}-2/\sqrt{E})}{\sqrt{\pi\sqrt{3}}}}\bigg[1- \displaystyle{\frac{4\sqrt{3}}{\pi\sqrt{E}}}\bigg]^{-1/2}.
\end{equation}
Let $I$ denote the spectral interval $[E_0,E_1]$, then 
\begin{equation*}
P[\eta, I] = \displaystyle{\frac{1}{N_{I}}} \sum_{E_{j} \in I} \delta (\eta - \eta_{j}) \approx \displaystyle{\frac{1}{N_{I}}} \int\limits_{E_0}^{E_1} \int\limits_{0}^{\pi/2} \delta [\eta - \eta_{j}]\, \mathrm dE \,\mathrm d\theta,
\end{equation*}
where $\displaystyle{\\N_I = \displaystyle{\frac{\sqrt{3}\pi}{16}} \bigg[ \big(E_1-E_0\big) - \displaystyle{\frac{4\sqrt{3}}{\pi}}\big(\sqrt{E_1}-\sqrt{E_0}\big)\bigg]}$.
\\Introducing the variables $x = \sqrt{\displaystyle{\frac{E}{E_0}}},\, g = \sqrt{\displaystyle{\frac{E_1}{E_0}}} \mbox { and } \epsilon = \displaystyle{\frac{4\sqrt{3}}{\pi\sqrt{E_0}}}$, we obtain
\begin{equation}
P(\eta, I) = \displaystyle{\frac{2E_0}{N_{I}}} \int\limits_{1}^{g} \int\limits_{0}^{\pi/2}x\, \delta \bigg[\eta - \displaystyle{\frac{8(\cos{\theta}-2/x \sqrt{E_0} )}{\sqrt{\pi\sqrt{3}}\,(1-\epsilon/x)^{1/2}}}\bigg]\, \mathrm dx \,\mathrm d\theta.
\end{equation}
Substituting $\cos{\theta} = y$, the distribution is reformulated as 
\begin{eqnarray*}
P(\eta, I) &=& \displaystyle{\frac{2E_0}{N_{I}}} \int\limits_{1}^{g}\,x \int\limits_{0}^{1}\,\displaystyle{\frac{1}{\sqrt{1-y^2}}}\, \delta \bigg[\eta - \displaystyle{\frac{8(y-2/x \sqrt{E_0} )}{\sqrt{\pi\sqrt{3}}\,(1-\epsilon/x)^{1/2}}}\bigg]\, \mathrm dx \,\mathrm dy \nonumber \\
&=& \displaystyle{\frac{E_0\sqrt{\pi\sqrt{3}}}{4N_I} \int\limits_{1}^{g}\,x(1-\epsilon/x)^{1/2} \int\limits_{0}^{1}}\,\displaystyle{\frac{1}{\sqrt{1-y^2}}}\, \delta \bigg[y - \displaystyle{\frac{16+\sqrt{\pi E_0\sqrt{3}}\:\eta\:(x^2-\epsilon x)^{1/2}}{8x\sqrt{E_0}}}\bigg]\, \mathrm dx \,\mathrm dy \nonumber \\
&=& \displaystyle{\frac{E_0\sqrt{\pi\sqrt{3}}}{4N_I}} \int\limits_{1}^{l} \displaystyle{\frac{x \, (1-\epsilon/x)^{1/2}}{\sqrt{1-\{f(x)\}^2}}}\, \mathrm dx, \mbox { where } f(x) = \displaystyle{\frac{16+\sqrt{\pi E_0\sqrt{3}}\:\eta\:(x^2-\epsilon x)^{1/2}}{8x\sqrt{E_0}}} \nonumber \\
\end{eqnarray*}
and  $l = \begin{cases}
g, & \hspace{1cm}\text{if }\hspace{1cm} \eta < \varphi_1,\newline \newline
\\ 
min\big[g, X_{max}\big], & \hspace{1cm} \text{if }\hspace{1cm} \eta > \varphi_1,
\end{cases}$\newline \newline
\\with $\varphi_1$ being defined as\newline
\\$\varphi_1 = \begin{cases}
\displaystyle{\frac{8}{3}}\sqrt{6-\sqrt{3}\pi}, \hspace{2.3cm}\text{if }\hspace{1cm} 0<\sqrt{E_0} \le \displaystyle{\frac{2\sqrt{3}}{\pi-\sqrt{3}}},\newline \newline
\\
\displaystyle{\frac{8}{3^{1/4}}}\displaystyle{\frac{\sqrt{E_0}-2}{\sqrt{ \pi E_0-4\sqrt {3E_0}}}}, \hspace{0.9cm} \text{otherwise}.
\end{cases}$

It is to be noted that $P(\eta, I) = 0$ for all $\eta > \varphi_2$, where $\varphi_2$ is specified as\newline
\\$\varphi_2 = \begin{cases}
\displaystyle{\frac{8}{3^{1/4}}}\displaystyle{\frac{\sqrt{E_0}-2}{\sqrt{ \pi E_0-4\sqrt {3E_0}}}}, \hspace{1cm}\text{if }\hspace{1cm} \displaystyle{\frac{4\sqrt{3}}{\pi}}<\sqrt{E_0} \le \displaystyle{\frac{\pi}{\pi-\sqrt{3}}},\newline \newline
\\
\displaystyle{\frac{8}{3^{1/4}\sqrt{\pi}}}, \hspace{3.2cm}\text{if }\hspace{1cm} \sqrt{E_0} >{\frac{\pi}{\pi-\sqrt{3}}}.\newline \newline
\end{cases}$\newline
\\However, there does not exist any such maximum value of $\eta$ beyond which $P(\eta, I) = 0$ in the case when $\sqrt{E_0} \le \displaystyle{\frac{4\sqrt{3}}{\pi}}$. In the definition of $l$ stated above, $X_{max}$ is the maximum permissible value of $x$ as specified by the inequality
\begin{equation*} 
0<\varphi_1 \le \displaystyle{\frac{8x\sqrt{E_0} - 16}{\sqrt{\pi E_0 \sqrt 3 (x^2-\epsilon x) }}} \le \varphi_2.
\end{equation*}
We evaluate $X_{max}$ for the most general case considering sufficiently excited states such that $E_0>\bigg(\displaystyle{\frac{2\sqrt 3}{\pi-\sqrt{3}}}\bigg)^2$. The integral for $P(\eta,I)$ can now be easily evaluated numerically for any given value of the parameter $E_0$ to obtain the distribution function for the normalised number of boundary intersections. The asymptotic distribution function has been plotted for $x = \sqrt{2}$ and $E_0 = 2000^2$ in Figure \ref{PEta} wherein the analytical and experimental results have been compared. Its inability to reproduce all the structures observed in the numerical data can be accounted for by recognising that the asymptotic result derives the distribution of $\tilde{\nu}_j$ in the non-tiling situation when at least one of $m+n$ and $m-n+1$ is not divisible by 3 as per \eqref{eq: bdry_spl}. However, the discrepancies with the actual distribution arise due to the fact that the spectral interval $[2000^2, 2\times 2000^2]$ includes multiple eigenstates which are in violation of the aforementioned condition as also wavefunctions corresponding to the tiling cases  --- the contributions from these additional states lead to the formation of the distinct maxima observed. In the non-tiling case considered for simplicity, the primary success of the analytical consideration stems from the theoretical prediction of the existence of a system dependent parameter $\eta ' = \varphi_2$, the maximum value of the normalised number of boundary intersections, such that $P(\eta) = 0\: \forall \: \eta > \eta '$. As can be observed from the graph, the value of $\varphi_2$ from the above-mentioned expression for $E_0 = 2000^2$ is 3.43, in close agreement with the numerical result, which indicates that the distribution $P(\eta ) $ is nearly zero beyond approximately 3.35.   

\subsection{The normalised number of nodal domains}
\begin{figure} [H]
\begin{flushleft}
\scalebox{0.478}{\includegraphics{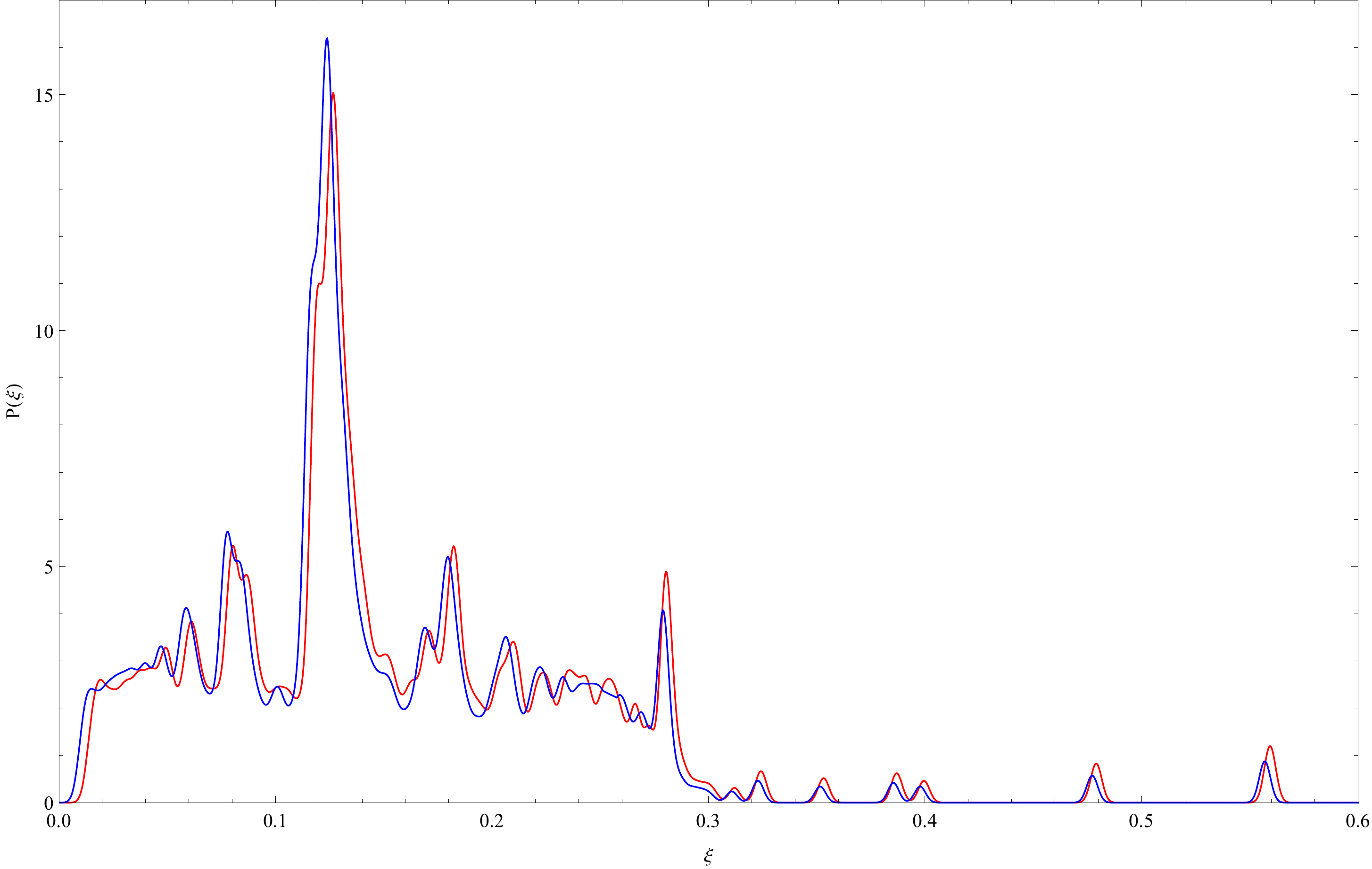}}
\end{flushleft}
\caption{A plot of the distribution function $P[\xi, I_{\lambda}(E)]$ of $\xi$, the normalised nodal domain number associated with each wavefunction, for the equilateral triangle billiard in the spectral intervals $[10000, \,20000]$ (red) and  $[20000, \,40000]$ (blue). The shift in the curves while preserving the basic form suggests the limiting behaviour as $E \rightarrow \infty$.}
\label{fig: pxi}
\end{figure} 
Each wavefunction $\psi_{m,n}$ may be associated with the normalised nodal domain number $\displaystyle \xi_{j} = \frac{\nu_{j}}{j}, 0 < \xi_j <1$. Taking the spectral interval $I_\lambda(E)$ to be as defined previously for \eqref{eq: eta}, the $\xi$ distribution for the interval may be similarly defined as
\begin{equation}
P[\xi, I_{\lambda}(E)] = \frac{1}{N_{I}} \sum_{E_{j} \in I_{\lambda}(E)} \delta \bigg(\xi - \frac{\nu_{j}}{j} \bigg) = \frac{1}{N_{I}} \sum_{E_{j} \in I_{\lambda}(E)} \delta (\xi -
\xi_{j}).
\label{eq: xi}
\end{equation}
The limiting distribution of the parameter $\xi$ is consequently obtained from \eqref{eq: xi} as
\begin{equation}
P(\xi) = \lim_{E \rightarrow \infty} P[\xi, I_{\lambda}(E)].
\end{equation}
For computational convenience, the Dirac delta function is well approximated by the Gaussian representation defined by
\begin{equation}
\delta(x) = \lim_{\sigma \rightarrow 0} \frac{e^{\frac{-x^2}{2 \sigma^2}}}{\sqrt{2 \pi} \sigma}.
\end{equation}
\begin{figure} [H]
\begin{flushleft}
\scalebox{0.75}{\includegraphics{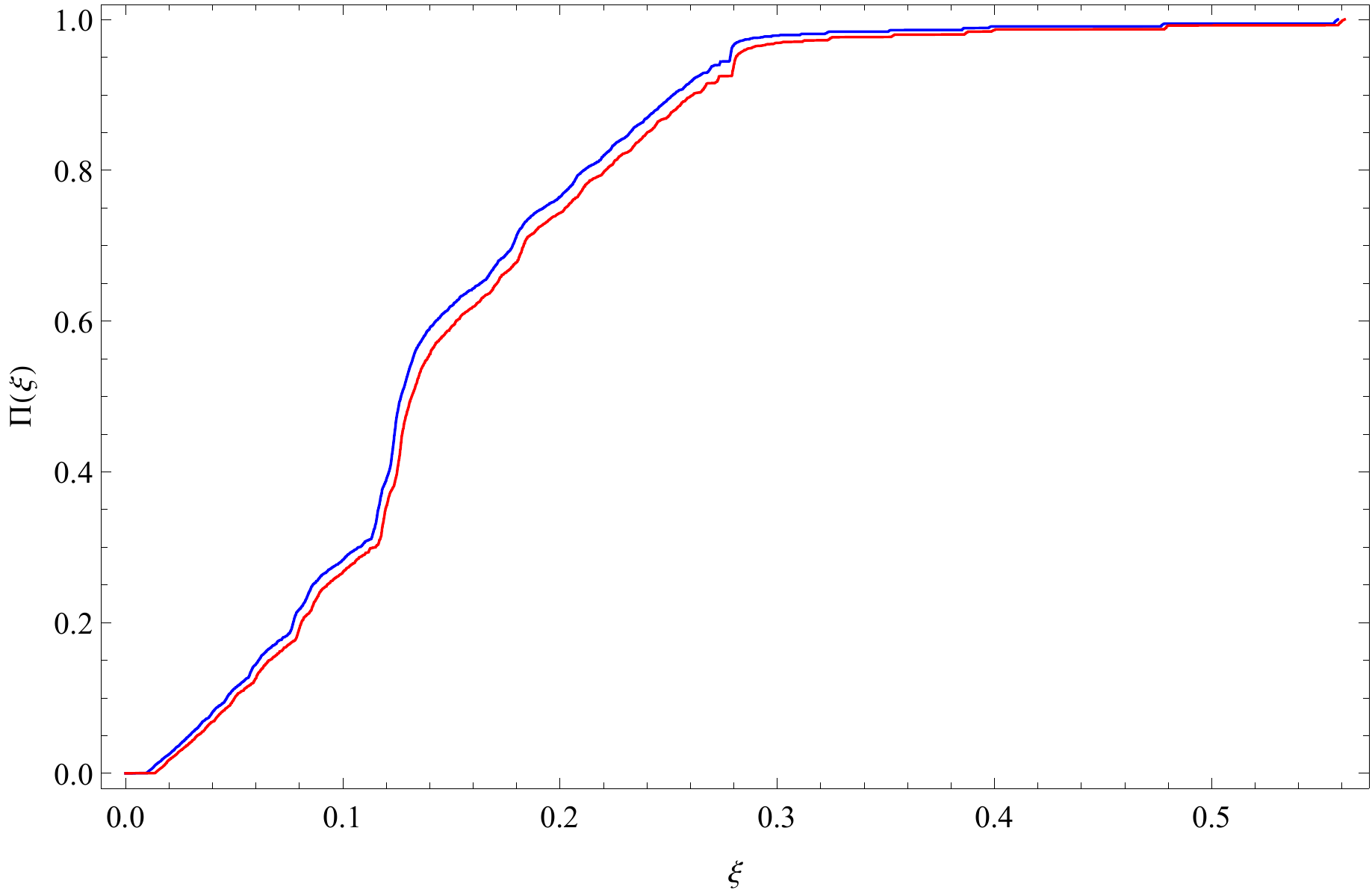}}
\end{flushleft}
\caption{ The integrated $\xi$ distribution for the equilateral triangle billiards, $\Pi\:[\xi, I_{\lambda}(E)]$ shown for all the wavefunctions in the spectral intervals $[10000, \,20000]$ (red) and  $[20000, \,40000]$ (blue), where $\xi$ represents the normalised number of nodal domains. The structural nature of both the curves is nearly identical. The distribution rises sharply at the same value of $\xi$ for which the principal peak is observed in the plot of $P(\xi)$ against $\xi$. }
\end{figure} 
The distribution $P[\xi, I_{\lambda}(E)]$ is shown in Figure \ref{fig: pxi} for all the eigenfunctions of the symmetric mode in the intervals $[10000, \,20000]$ and  $[20000, \,40000]$ with $\lambda = 1$ with the analysis being extended over $3014$ and $6028$ wavefunctions respectively. 
As observed for the boundary intersections count, the figure resembles neither the distribution for separable billards nor that of chaotic billiards. An extremely sharp fundamental peak is observed in the plot, which is flanked by three smaller but prominent secondary peaks. As the value of $E$ for the interval $I_1(E)$ increases, the entire distribution, including all the peaks, shifts towards the left while the peaks become narrower and sharper. Nevertheless, the inherent form or the basic structure of the distribution remains nearly unchanged with the variation in $E$. The translation of the peaks provides an insightful hint towards the asymptotic behaviour of  $P[\xi, I_{\lambda}(E)]$. However, it remains insufficient to conclude about the nature or existence of the limiting distribution, if any. Further attempts to obtain an analytical expression for this limiting distribution could proceed as per the methods outlined by \cite{karageorge}. The corresponding integrated distribution for the system is denoted by
\begin{equation}
\Pi\:[\xi, I_{\lambda}(E)] = \int_{0}^{\xi} \, P[x, I_{\lambda}(E)]\, \mathrm dx.
\end{equation}
It has been shown by Pleijel \cite{pleijel} that
\begin{equation}
\overline{\lim_{j \rightarrow \infty}}\: \frac {\nu_{j}}{j}\: \le \:\bigg(\frac{2}{j_{0}}\bigg)^2 \approx\: 0.691,
\end{equation}
where $j_{0}$ is the first zero of the Bessel function $J_0(.)$. It is evident from Figure \ref{fig: pxi} that the wavefunctions for the equilateral triangle respect the bound imposed by the aforementioned value of the limit superior since $P(\xi) \approx\: 0\:\, \forall \:\, \xi\: > 0.58$.

\subsection{The signed area distribution}

\begin{figure} [H]
\begin{center}
\subfloat[]{\includegraphics[scale = 0.52]{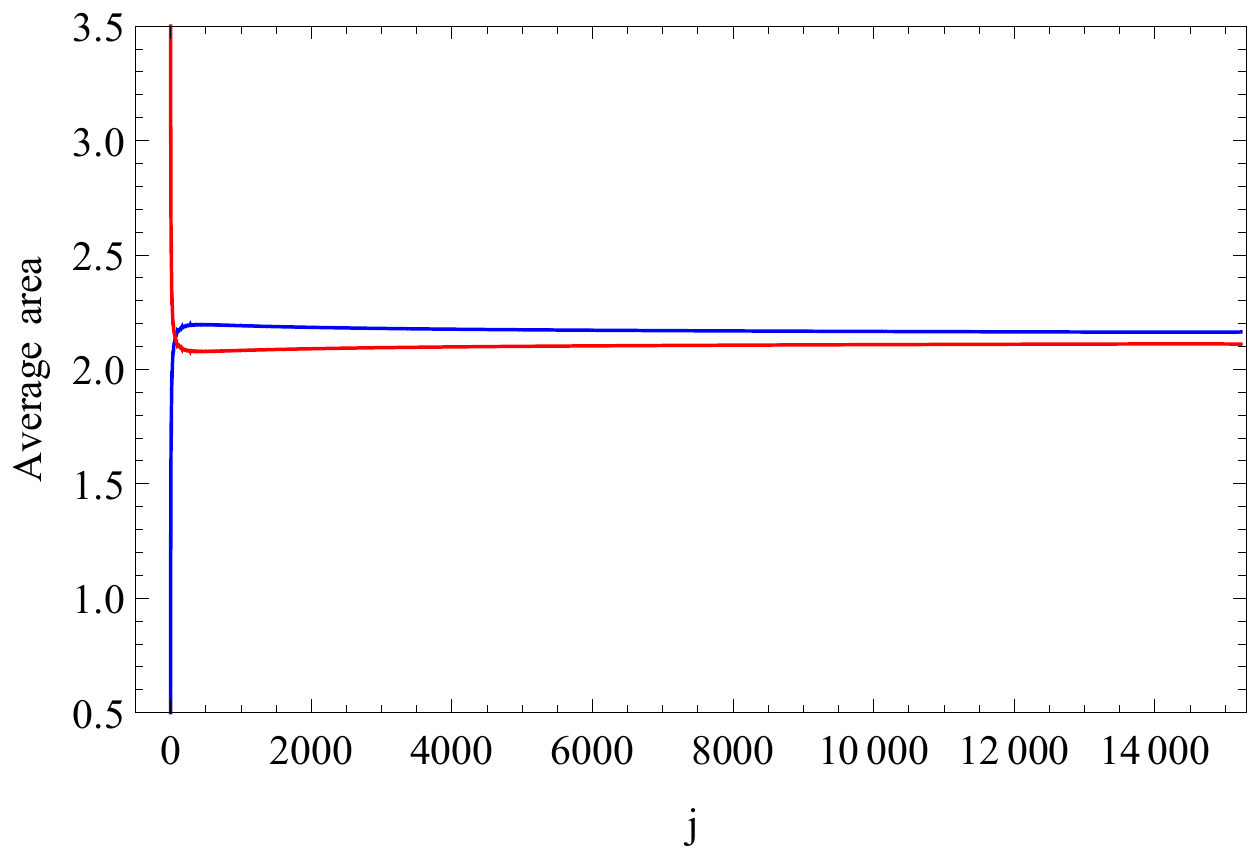}}
\quad
\subfloat[]{\includegraphics[scale = 0.545]{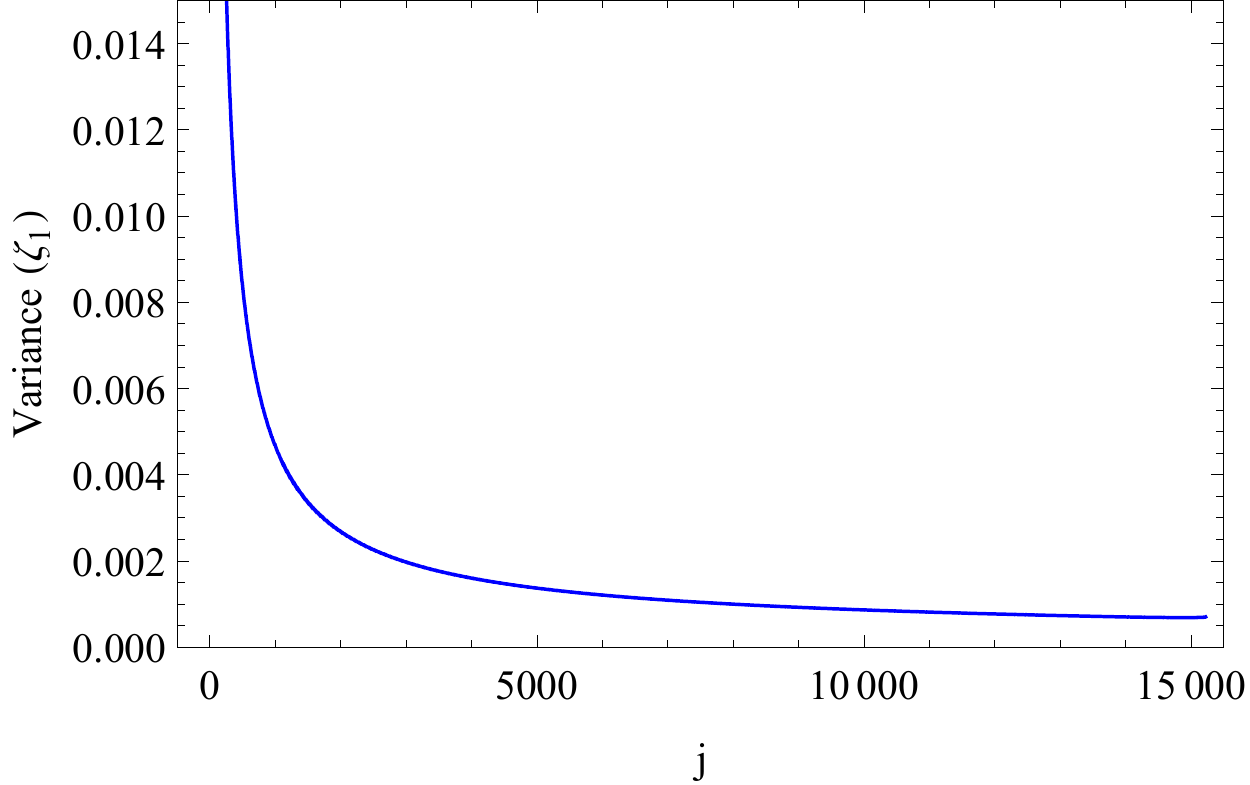}}
\end{center}
\caption{(a) The running averages of the total positive (blue) and negative (red) areas of the domains over the first 7650 eigenfunctions as a function of the index $j$ which shows the asymptotic convergence of the two curves to a common value $\displaystyle = \frac{|\Omega|}{2}$ as $j \rightarrow \infty$. The convergence of the averages suggests that it is likely that the area of the positive and negative domains equalise in the limit of extremely large values of $j$.(b) The normalised signed area variance for the equilateral triangle billiards (running averages of 7650 eigenfunctions considered). }
\label{area}
\end{figure}

\begin{figure} [H]
\begin{center}
\scalebox{0.66}{\includegraphics{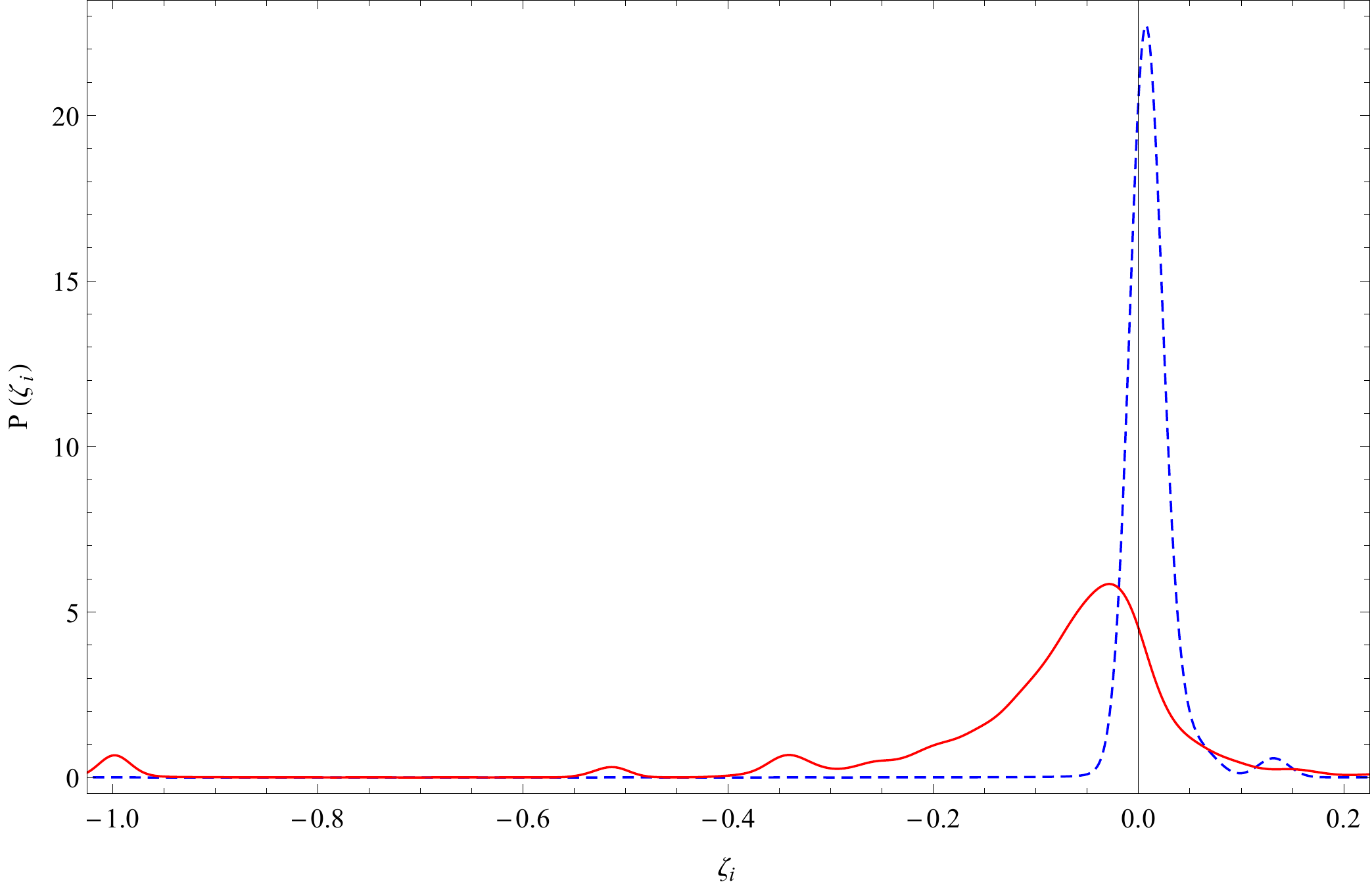}}
\end{center}
\caption{The probability distributions of the normalised variables $\zeta_1$ and $\zeta_2$, depicted as the dashed and solid curves respectively. Interestingly, both the curves are off-centred and neither is symmetric about zero. The total number of the negative domains is likely to slightly exceed that of the positive, but the reverse holds for the respective total areas of the positive and negative domains.}
\label{area2}
\end{figure}

The area of a nodal domain of the $j$-th eigenfunction is bounded by the relation \cite{krahn} 
\begin{equation}
\label{eq:area}
{\frac{\pi\, (j_0)^2}{k_j^2}\: \leq \:\cal A}(\psi _j),
\end{equation} 
where $j_0$ is the first zero of the zeroth order Bessel function $J_0$.  
Let $|\Omega (\psi)|_{+}$ and $|\Omega (\psi)|_{-}$ represent the total area of the domain where the wavefunction is positive and negative respectively. Furthermore, $\nu_+$ and $\nu_-$ are used to denote the total number of positive and negative domains of the wavefunction.  Figure \ref{area} (a) presents the running averages of the positive and negative areas as a function of the index $j$.  For the initial values of $j$,  $\langle |\Omega (\psi)|_{+} \rangle \ne \langle|\Omega (\psi)|_{-}\rangle$, as expected, but the curves progressively approach a common value and seemingly tend to converge as $j \rightarrow \infty$. Hence, it is probable that for sufficiently large values of $j,\, \displaystyle \langle |\Omega (\psi)|_{+} \rangle = \langle|\Omega (\psi)|_{-}\rangle = \frac{|\Omega|}{2}$, where $|\Omega|$ is the total area of the billiards. \\
\\Let $\zeta_1$ and $\zeta_2$ be two scaled variables defined as $\displaystyle \zeta_{1} = \frac {|\Omega (\psi)|_{+} - |\Omega (\psi)|_{-}}{|\Omega|}$ and $\displaystyle \zeta_{2} = \frac {\nu_{+} - \nu _{-}}{\nu}$. The variance of the parameter $\zeta_1$ defines the normalised signed area variance of the system and is shown in Figure \ref{area} (b). The probability distributions of $\zeta_{1} \mbox{ and }\zeta_{2}$ are shown in Figure \ref{area2}. 

The distribution function for $\zeta_{1}$ peaks at a slightly positive value while that of $\zeta_{2}$ presents the principal peak at a negative value along with three smaller but significant peaks further towards the negative side. The distribution function of $\zeta_2$ shows that the number of negative domains is likely to be greater than the number of positive domains for a given state, i.e. $\zeta_2 < 0.$ However, it can also be seen that the total area of the positive domains is likely to be slightly greater than the total area of the negative domains, i.e. $\zeta_1 > 0$ for a particular state. 

\subsection{The cumulative nodal loop count}

The statistics of the number of boundary intersections have already been analysed extensively in Section 2.2. The present discussion focusses on the nodal loop count. The two continuous cumulative counting functions introduced by \cite{aronovitch} are defined as
\begin{eqnarray*}
C(N) &:=& \sum_{j = 1}^{\lfloor N \rfloor} l_{j}
\\V(k) &:=& \sum_{j = 1}^{\infty}l_j\Theta (k - k_j)
\end{eqnarray*}
where the nodal loop count of the $j$th eigenfunction is denoted by $l_{j}$ and $k_j = \sqrt{E_j}$. The nodal counting function can be expressed as a trace formula for surfaces of revolution \cite{count} which comprises of a smooth (Weyl-like) part determined by the geometric parameters of the domain, and an oscillatory component dependent on the lengths of the classical periodic orbits on the domain and their actions. This intricate connection of the classical periodic orbits with the sequences of nodal counts rationalises a similar investigation of the counting functions defined previously for the equilateral triangle billiards. When examined numerically, both the functions $C(N)$ and $V(k)$ are noticed to have a reasonably well-defined smooth term and an oscillating term. Similar to the case of separable billiards, the smooth part of $C(N)$ is approximately $O(N^2)$ while that of $V(k)$ is very nearly $O(k^4)$. The oscillatory components of both $C(N)$ and $V(k)$ are easily obtained by subtracting out the smooth part, after interpolation, from the tables of evaluated values of the corresponding functions. The contribution of the periodic orbits is ascertianed by considering the Fourier transform of $C_{osc}(N)$ and $V_{osc}(k)$. 
\begin{figure} [H]
\begin{center}
\includegraphics[scale = 0.475, trim = 0cm 3cm 0cm 0cm]{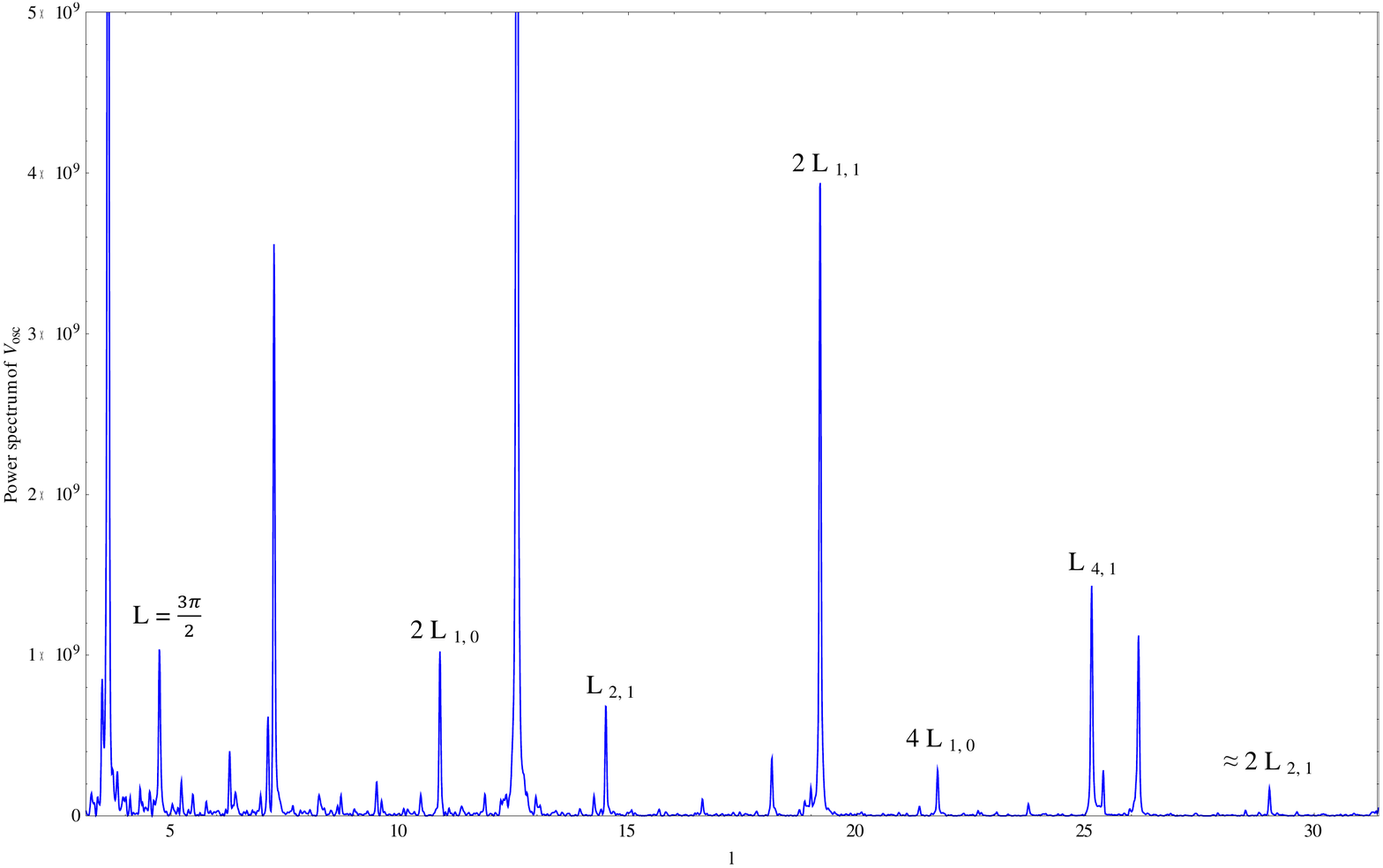}
\end{center}
\caption{The power spectrum obtained on Fourier transforming the oscillatory part of the cumulative counting function, $V(k)$, for the nodal loops. The lengths of some of the prominent periodic orbits in the interval $L_{p,q} \in [\pi,10\pi]$ are marked on the $l$ axis.}
\label{PO2}
\end{figure} 
The Fourier transform of $C_{osc}(N)$ is carried out not with respect to $N$ but rather, with respect to the variable $\displaystyle c = \sqrt{\frac{4\pi}{{\cal A}}N} = 4\sqrt{\frac{N}{\sqrt{3}\pi}}$, following \cite{aronovitch}.  
The lengths of the classical periodic orbits on an equilateral triangular domain, having each side of length $a$, are given by \cite{{brack}, {robinett}} as $\displaystyle L_{p,q} = a\sqrt{3(p^2+pq+q^2)}$, where $(p,q) \in \mathbb{Z}^2\backslash (0,0)$. Such a closed orbit of length $L_{p,q}$ makes an initial angle of $\displaystyle \tan ^{-1}\bigg(\frac{p-q}{(p+q)\sqrt{3}}\bigg)$ with the horizontal.
\begin{figure} [H]
\begin{center}
\includegraphics[scale = 0.485, trim = 0cm 5.0cm 0cm 0cm]{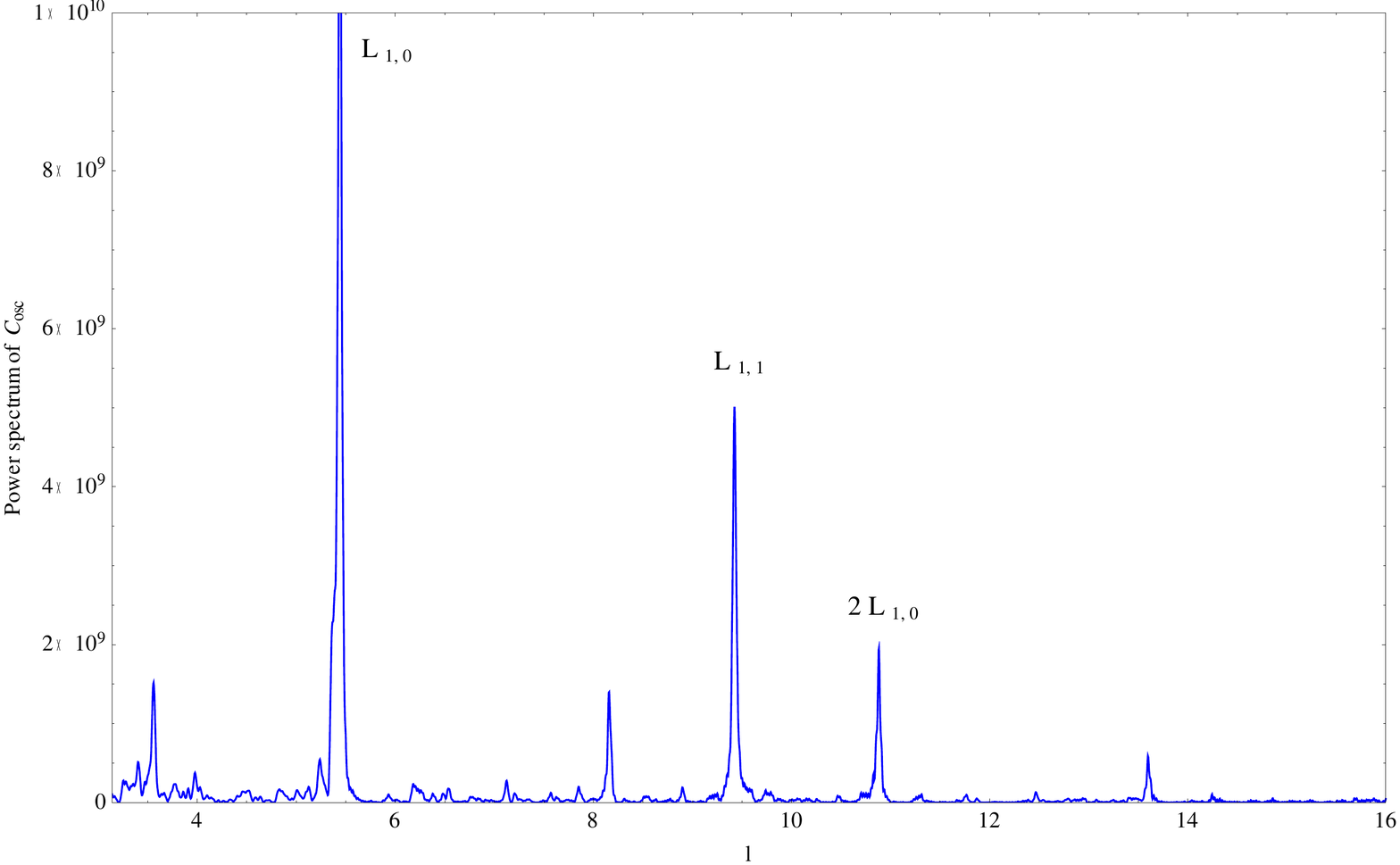}
\end{center}
\end{figure}
\begin{figure} [H]
\begin{center}
\includegraphics[scale = 0.475, trim = 0cm 3cm 0cm 0cm]{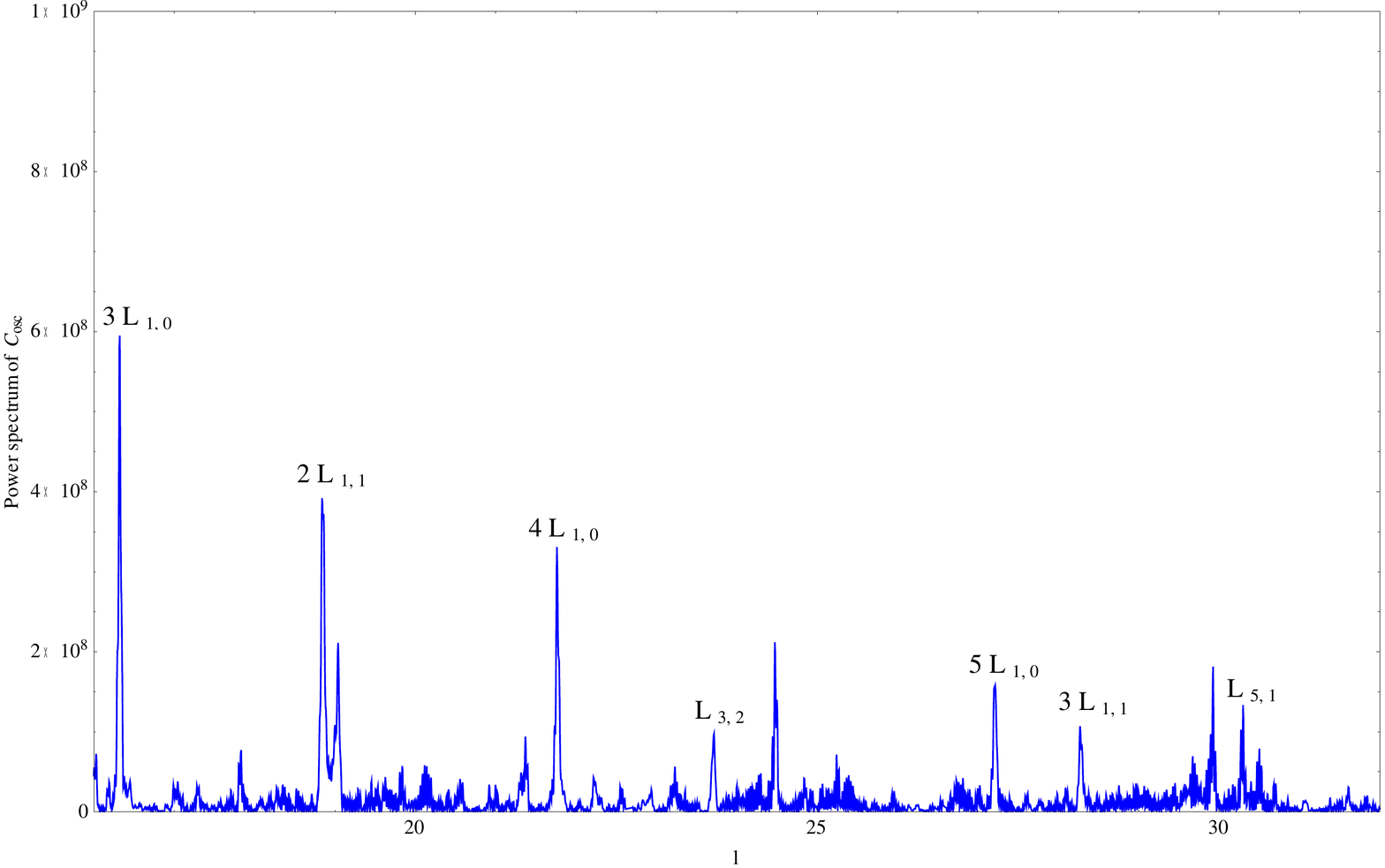}
\end{center}
\caption{The power spectrum obtained on Fourier transforming the oscillatory part of $C(N)$ with respect to the scaled variable $c$. The lengths of some of the prominent periodic orbits in the interval $L_{p,q} \in [\pi,10\pi]$ are marked on the $l$ axis.}
\label{PO1}
\end{figure} 

A linear combination of peaks is observed on inspecting the Fourier transform of $V_{osc}(k)$.  The prominent contributions of the periodic orbits belonging to certain important families are noticed in both Figures \ref{PO2} and \ref{PO1} and these have been enumerated in the following discussion.
Moreover, the power spectrum of $V_{osc}(k)$ also shows a significant peak at a length of $\displaystyle \frac{3\pi}{2}$. This corresponds to an isolated orbit, the trajectory of which is simply the pedal triangle of the equilateral triangle. The power spectrum also shows characteristic peaks at values of $l = 2z\pi, z\in \mathbb{N}$, although there exists no closed trajectories of such lengths, thus implying the presence of additional factors in the billiard's dynamics.\newline
Examination of the power spectrum of $C_{osc}(N)$ also shows a number of distinct peaks of similar nature, the positions of some of which exactly correspond to the lengths of certain periodic orbits. \cite{robinett,svl} tabulates the path lengths for the closed orbits (and recurrences) on the equilateral triangle for which $\displaystyle \frac{L_{p,q}}{a} < 20$. \\
Although several periodic orbits corresponding to different values of the integers $p$ and $q$ are shown in Figure \ref{PO1}, there is a clear demarcation of two important families.
\begin{enumerate}
\item The trajectories, which start off parallel to the horizontal and bounce from each side of the triangle twice to ensure closure of the path. The length of any orbit of the family is naturally $L = 3z\pi, z\in \mathbb{N}$. 
\item The closed paths that traverse a length of $\sqrt{3}\pi$ in each period and start off at an initial angle of $30^{\circ}$ with respect to the horizontal.
\end{enumerate}
The fingerprints of the classical periodic orbits noted in the Fourier transforms of $C(N)$ and $V(k)$ strongly hint at the possibility of the existence of a trace formula for both the quantities.

\section{Summary and discussion}
Counting and classification of patterns is an important and difficult task. Here we have tried to understand the spatial patterns generated by the signs of the eigenfunctions of the equilateral triangle billiard. This is a nonseparable, integrable system for which we know the solutions of the Helmholtz equation since the times of Lam\`{e}. The approach is phenomenological insofar as it is guided by the numerical coding of the patterns but we have nevertheless obtained equations from the plethora of data, which we have been able to solve. The solutions are verified over a large number of eigenfunctions. The number of nodal domains, boundary intersections, nodal loops and the signed area distribution are all studied. For each of these quantities, we have obtained expressions and given convincing evidence of the results in terms of analytical and numerical forms.  

We believe the connections we have found between the arrangement and `evolution' of the triplets (elaborated further in the Appendix) with the theory of groups is exciting. In a sense, this observation encodes the beauty of the nodal patterns in the form of triplets that provide a link with the underlying symmetry group. The periodic recurrence of these indices (fractions) was unanticipated as was the belief that they would encode the patterns analysed in this article. The connection of the statistical quantities with periodic orbits of the billiard, which was expected owing to earlier works, has been verified and extended for the equilateral triangle domain herein. Several results indicate suggestive comparisons and departure from the results known for separable and chaotic billiards. 

We believe that the studies and results found here will pave the way for future work on non-separable and polygonal billiards. 
\newpage

\section*{Appendix A}
Two examples of the group-theoretic considerations of the nodal loops of the equilarateral triangle billiard, discussed in Section 2.4, are presented here for illustration and corroboration of the ideas introduced.
\begin{figure} [H]
\begin{center}
\subfloat[]{\includegraphics[scale = 0.55, trim = 0cm 7cm 0cm 0cm]{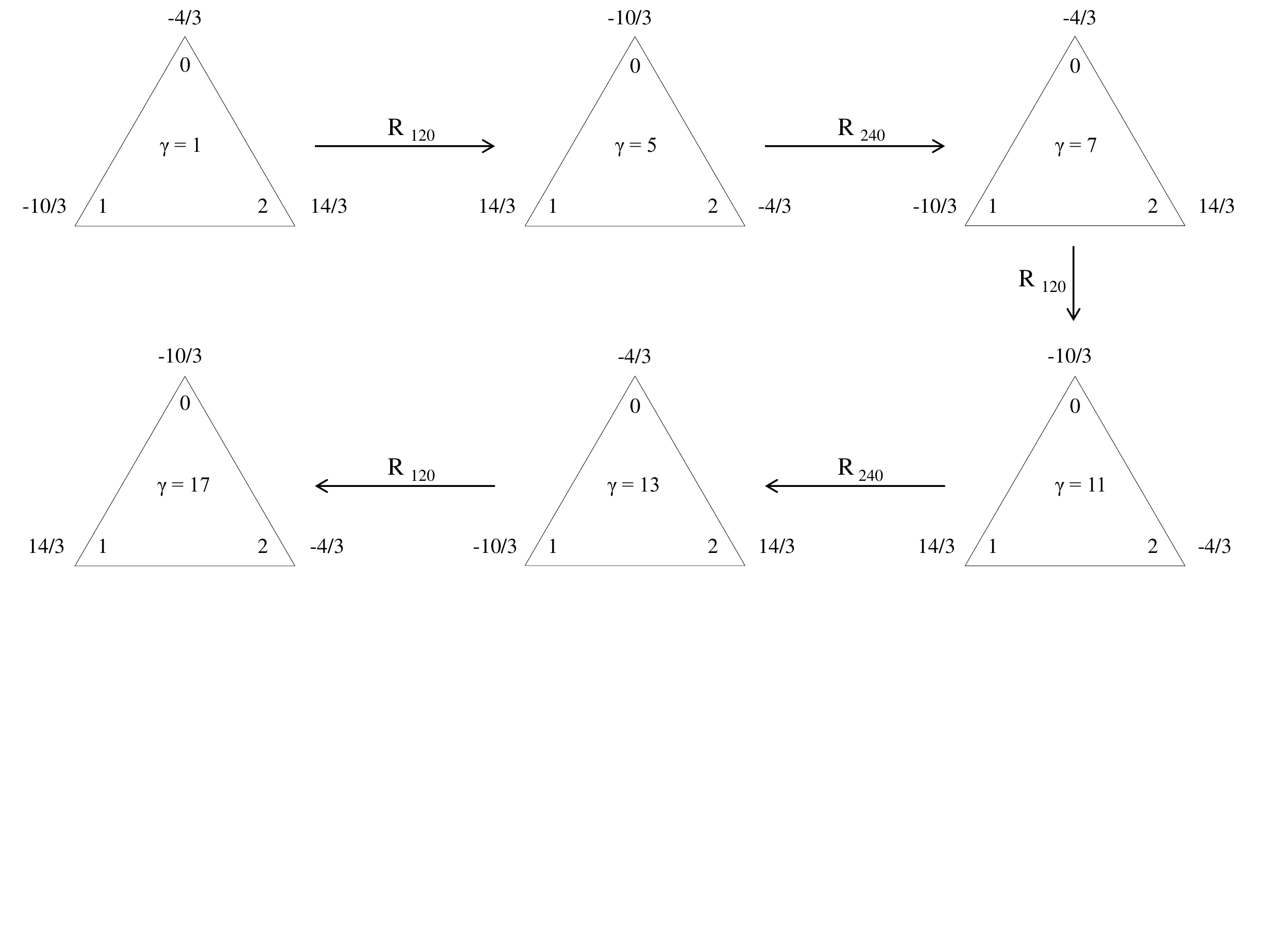}}
\quad
\\
\subfloat[]{\includegraphics[scale = 0.55, trim = 0cm 7cm 0cm -1.5cm]{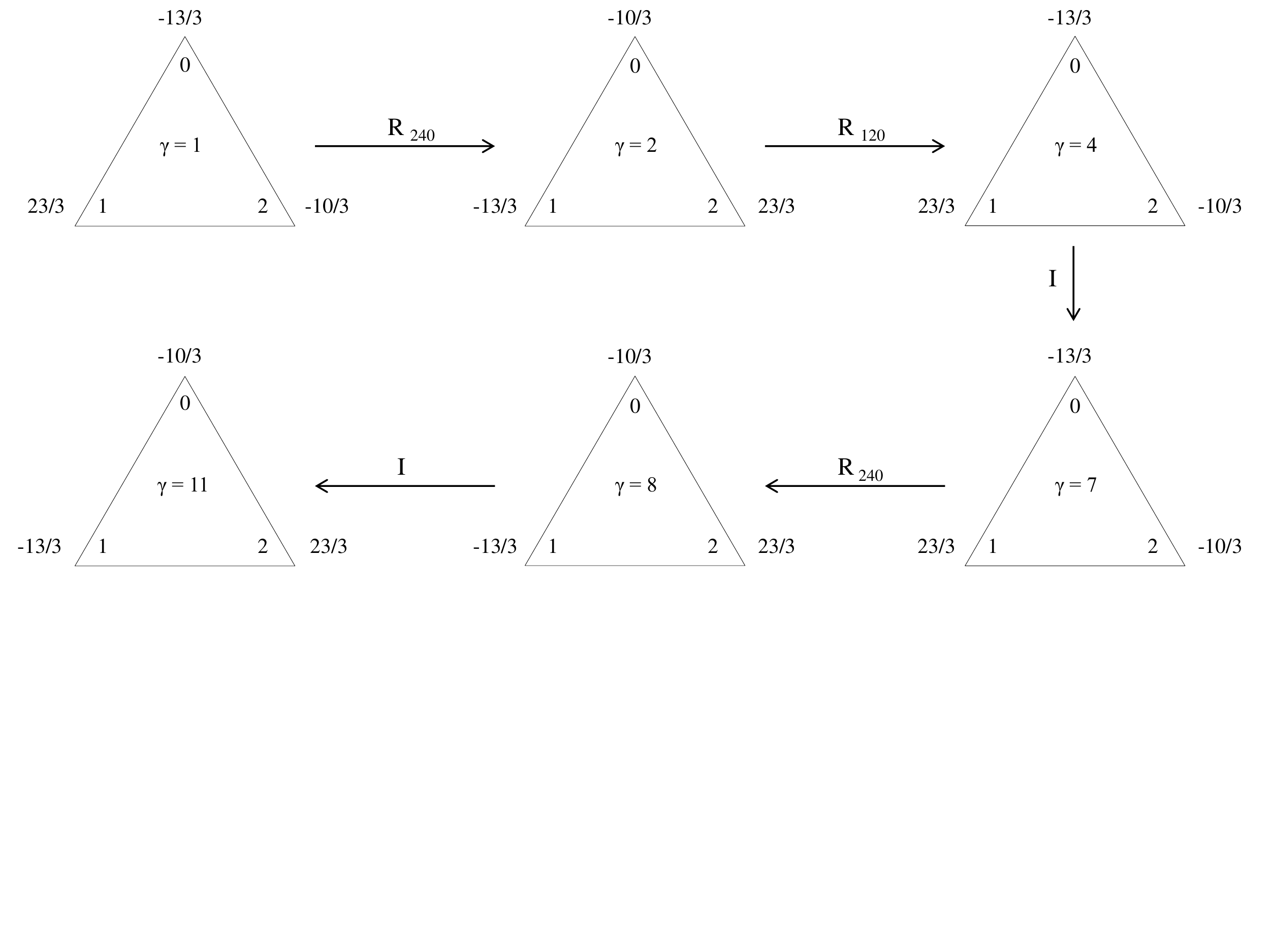}}
\end{center}
\caption*{Figure A: The permutation of the labels $\vartheta_{i}$, corresponding to change in $\gamma$, with respect to the vertices labelled by $m_{i}\mod3$, shown for (a) $n = 4, c_{I} = 1$ and (b) $n = 5, c_{I} = 2$. The rotation operation for each transformation is indicated. The composition law of the $C_3$ group is evident herein. The transformation from $\gamma \rightarrow \gamma +z,\, z\in \mathbb{N}$ is effected by $z$ succesive applications of the $R_{120}$ and $R_{240}$ operators in the first and second case respectively.}
\label{group}
\end{figure}
Proceeding with the definition of $\vartheta$ provided by \eqref{eq: omega}, it is observed that the set consists of only three possible values whenever all the quantum numbers $m$ considered are multiples of a certain fixed $\gamma \in \mathbb{N}$ and belong to the same equivalence class $c_I = m\mod n$. More importantly, the values of $\vartheta_{i}$ corresponding to $m$ and $m + \gamma n$ are the same (provided $c_I$ and $n$ remain constant), which shows the existence of a well-defined periodicity of $\gamma$ with respect to $m$. The set $\vartheta$ has been enumerated for different values of $n$ and $c_I$ and the three possible values in each case are listed below in the following table.\\
{\small
\begin{center}
\begin{longtable}{c c | c c c}
\hline
$n$ & $\displaystyle c_I$ & $\vartheta_{1}$ &  $\vartheta_{2}$ & $\vartheta_{3}$ \endhead \hline
3 & 1 & -2 & 1 & 1 \\
4 & 1 & $\displaystyle \frac{-10}{3}$ & $\displaystyle \frac{-4}{3}$ & $\displaystyle \frac{14}{3}$ \\ 
5 & 1 & $\displaystyle \frac{-28}{3}$ & $\displaystyle \frac{5}{3}$ & $\displaystyle \frac{23}{3}$ \\ 
5 & 2 & $\displaystyle \frac{-13}{3}$ & $\displaystyle \frac{-10}{3}$ & $\displaystyle \frac{23}{3}$ \\ 
6 & 1 & -8 & -2 & 10 \\
7 & 1 & $\displaystyle \frac{-28}{3}$ & $\displaystyle \frac{-25}{3}$ & $\displaystyle \frac{53}{3}$ \\ 
7 & 2 & $\displaystyle \frac{-46}{3}$ & $\displaystyle \frac{5}{3}$ & $\displaystyle \frac{41}{3}$ \\ 
7 & 3 & $\displaystyle \frac{-31}{3}$ & $\displaystyle \frac{8}{3}$ & $\displaystyle \frac{23}{3}$ \\ 
8 & 1 & $\displaystyle \frac{-58}{3}$ & $\displaystyle \frac{-16}{3}$ & $\displaystyle \frac{74}{3}$ \\ 
8 & 3 & $\displaystyle \frac{-40}{3}$ & $\displaystyle \frac{17}{3}$ & $\displaystyle \frac{23}{3}$ \\ 
9 & 1 & -20 & -11 & 31 \\
9 & 2 & -22 & -4 & 26 \\
9 & 4 & -22 & 8 & 14 \\
11 & 1 & $\displaystyle \frac{-106}{3}$ & $\displaystyle \frac{-55}{3}$ & $\displaystyle \frac{161}{3}$ \\ 
11 & 2 & $\displaystyle \frac{-91}{3}$ & $\displaystyle \frac{-52}{3}$ & $\displaystyle \frac{143}{3}$ \\ 
11 & 3 & $\displaystyle \frac{-91}{3}$ & $\displaystyle \frac{17}{3}$ & $\displaystyle \frac{74}{3}$ \\ 
11 & 4 & $\displaystyle \frac{-121}{3}$ & $\displaystyle \frac{29}{3}$ & $\displaystyle \frac{92}{3}$ \\ 
11 & 5 & $\displaystyle \frac{-85}{3}$ & $\displaystyle \frac{17}{3}$ & $\displaystyle \frac{68}{3}$ \\ 
13 & 1 & $\displaystyle \frac{-130}{3}$ & $\displaystyle \frac{-109}{3}$ & $\displaystyle \frac{239}{3}$ \\ 
\hline
13 & 2 & $\displaystyle \frac{-148}{3}$ & $\displaystyle \frac{-61}{3}$ & $\displaystyle \frac{209}{3}$ \\ 
13 & 3 & $\displaystyle \frac{-124}{3}$ & $\displaystyle \frac{-7}{3}$ & $\displaystyle \frac{131}{3}$ \\ 
13 & 4 & $\displaystyle \frac{-142}{3}$ & $\displaystyle \frac{-4}{3}$ & $\displaystyle \frac{146}{3}$ \\ 
13 & 5 & $\displaystyle \frac{-151}{3}$ & $\displaystyle \frac{35}{3}$ & $\displaystyle \frac{116}{3}$ \\ 
13 & 6 & $\displaystyle \frac{-118}{3}$ & $\displaystyle \frac{56}{3}$ & $\displaystyle \frac{62}{3}$ \\ 
15 & 1 & -62 & -47 & 109 \\
15 & 2 & -64 & -34& 98 \\
15 & 4 & -67 & -1 & 68 \\
15 & 7 & -55 & 23 & 32 \\
\hline
\caption{List of numerically evaluated values of $\vartheta_{1}, \vartheta_{2}$ and $\vartheta_{3}$ obtained for different values of the quantum number $n$ and the residue class $c_I$. It may be verified that $\textstyle \sum_{i = 1}^3 \vartheta_{i} = 0$. Hence, the corresponding values of $\theta_{i}$ are obtained using the relation $\displaystyle \theta_{i} = \vartheta_{i} + \frac{\gamma^2 \,n^2}{3} $.}
\end{longtable}
\end{center}
}
For observation of the underlying group-theoretic properties of the system, it is convenient to adopt the pictorial representation of an equilateral triangle, the vertices of which are marked with the three values of $m_{i}\mod3 = \{0, 1, 2\}.$ Such a mode of depiction is only for ease of visualisation (since there are only three distinct values in the set $\vartheta$) and is merely a representational means of understanding the symmetry relations. The vertex tagged with $m_{i}\mod3$ is assigned a label of $\vartheta_{i}$, where $\vartheta_{i}$ is the value obtained from the solution of \eqref{eq: omega} on substituting the corresponding $m_i$. The rearrangement of the labels of the vertices with respect to the variation of $\gamma$ is studied. Any such permutation can be conveniently represented by a specific rotation of the triangle constructed, as above. The examples shown here  serve to establish that these operations obey the composition law of the $C_3$ group whereby any combination of operations can be effectively implemented by a single transformation. Knowledge of the relative positions of the labels for any one value of $\gamma$ enables the extension of the diagram to higher values of $\gamma$ through succesive applications of a certain ascertainable transformation. Thereafter, calculation yields the number of nodal loops for higher values of the quantum number $m$ using the appropriate values of $\vartheta_{i}$ in \eqref{eq: omega} and hence, permits the determination of the number of nodal domains.

We would like to conclude by listing questions that are brought out of our study, and remain unanswered.

\begin{enumerate}

\item The form of the functions $F_1(c, n)$,  $F_2(c, n)$ appearing in (2.4) are not known. We have given the functional relations satisfied by them.

\item Existence of a limiting distribution of the mode number remains unproven. Fig. 3.1 suggests a rather complex form, although the study remains inconclusive. Numerical experiments for much higher-lying eigenfunctions need to be carried out. Of course, an analytic result will settle the matter.

\item Fig. 3.4 summarizes the numerical understanding of the area and number of nodal domains. It would be much more satisfying if one could present analytical results for these.

\item A lot of work has been done on eigenfunctions of simple pseudointegrable billiards (e.g. \cite{jain92,jain90}) like $\pi /3$-rhombus billiard. The connection of this system with the six-vertex model \cite{gaudin} and fluid mechanics makes it an exciting open problem. Since it is the simplest non-integrable, non-chaotic, non-separable billiard, any results on its nodal domains will be important. \newline

\end{enumerate}

\noindent
{\bf Acknowledgements}

The authors thank Ved V. Datar for an interesting discussion when they were in the process of interpreting the numerical results. This work was carried out while one of the authors (Rhine Samajdar) was visiting the other author (SRJ) at Bhabha Atomic Research Centre (BARC), Mumbai as a Summer Research Fellow under a programme conducted by The Indian Academy of Sciences, The Indian National Science Academy, and National Academy of Sciences, India. RS thanks the Academies and BARC for the support extended during the stay.


\begin{thebibliography}{99}
\bibitem{sravan} The graph-theoretic method was employed for the case of an equilateral triangle billiard by Nandhakumar Sravan. Although for some simple eigenfunctions, one can see its applicability, but it gets unwieldy quickly.
\end{thebibliography}
\end{document}